\documentclass{emulateapj}

\newcommand{\boldsymbol}[1]{\mbox{\boldmath{${#1}$}}}

\shorttitle{Evidence for dark matter contraction and a Salpeter IMF in a massive early-type galaxy}
\shortauthors{Sonnenfeld et al.}

\begin{document}

\title{Evidence for dark matter contraction and a Salpeter IMF in a massive early-type galaxy}

\author{A.~Sonnenfeld\altaffilmark{1}, T.~Treu\altaffilmark{1}, R.~Gavazzi\altaffilmark{2}, P.~J.~Marshall\altaffilmark{3}, M.~W.~Auger\altaffilmark{1,4}, S.~H.~Suyu\altaffilmark{1}, L.~V.~E.~Koopmans\altaffilmark{5}, A.~S.~Bolton\altaffilmark{6}}

\altaffiltext{1}{Department of Physics, University of California,
    Santa Barbara, CA 93106}
\altaffiltext{2}{Institut d'Astrophysique de Paris, UMR7095 CNRS - Universit\'{e} Pierre et
Marie Curie, 98bis bd Arago, 75014 Paris, France}

\altaffiltext{3}{Department of Physics, University of Oxford, Keble Road, Oxford, OX1 3RH, UK}
\altaffiltext{4}{Institute of Astronomy, University of Cambridge, Madingley Road, Cambridge CB3 0HA, UK}
\altaffiltext{5}{Kapteyn Institute, University of Groningen, PO Box 800, 9700 AV Groningen, the Netherlands}
\altaffiltext{6}{Department of Physics and Astronomy, University of Utah, Salt Lake City, UT 84112}

\email{sonnen@physics.ucsb.edu}

\begin{abstract}
Stars and dark matter account for most of the mass of early-type
galaxies, but uncertainties in the stellar population and the dark matter
profile make it challenging to distinguish between the
two components.  Nevertheless, precise observations of stellar
and dark matter are extremely valuable for testing the many
models of structure formation and evolution.  We present a
measurement of the stellar mass and inner slope of the dark matter
halo of a massive early-type galaxy at $z=0.222$.  The galaxy is the
foreground deflector of the double Einstein ring gravitational lens
system SDSSJ0946+1006, also known as the ``Jackpot''.  By combining
the tools of lensing and dynamics we first constrain the mean slope
of the total mass density profile ($\rho_{\rm{tot}}\propto
r^{-\gamma'}$) within the radius of the outer ring to be $\gamma' =
1.98\pm0.02\pm0.01$.  Then we obtain a bulge-halo decomposition,
assuming a power-law form for the dark matter halo.  Our analysis
yields $\gamma_{\rm{DM}} = 1.7\pm0.2$ for the inner
slope of the dark matter profile, in agreement with theoretical
findings on the distribution of dark matter in ellipticals, and a
stellar mass from lensing and dynamics $M_*^{\rm{LD}} =
5.5_{-1.3}^{+0.4}\times10^{11}M_\Sun$.  By comparing this measurement
with stellar masses inferred from stellar population synthesis fitting
we find that a Salpeter IMF provides a good description of the stellar
population of the lens
while the probability of the IMF being heavier than Chabrier is 95\%.
Our data suggest that growth by accretion of small systems from a compact red
nugget is a plausible formation scenario for this object.
\end{abstract}


\section{Introduction}

What are the main physical processes that shape early-type galaxies (ETGs)? 
Understanding the formation and evolution of early-type galaxies is a fundamental piece in the cosmological puzzle.
Any model that aims at providing a description of the Universe as a whole must be able to reproduce the observed characteristics of these objects.

ETGs are observed to hold tight scaling relations, such as the Fundamental Plane \citep{Dressler,DjorDavis} and the correlation between the mass of the central black hole and global galactic properties \citep{Ferrarese,Gebhardt,Marconi}.
The inner density profile of their total mass is measured to be very close to isothermal, in the so-called ``bulge-halo conspiracy'' \citep{TreuKoop,Koop2006,Koo++09,PaperX}.
Moreover, they appear to undergo a significant evolution in size, from being very compact in the early ($z\sim2$) Universe to the more diffuse objects that we observe at more recent times \citep{vanderWel2008,vanDokkum,Newman2010}.
Finally, studies of massive ETGs seem to favor a heavier stellar initial mass function \citep[IMF;][]{Grillo2009,TreuIMF,v+C10,v+C11,Auger2010L,Chiara} than for spiral galaxies \citep{B+d01,Dut++11,Suy++11}.

These characteristics should reflect a common mechanism that drives ETGs towards the tight relations observed at low redshifts ($z \lesssim 0.3$) during their formation and evolution.
Mergers with other galaxies are likely to be one of the key processes in the history of ETGs.
Mergers are believed to be at the basis of the formation \citep{Kormendy,Hernquist92,Shier}, to be involved in the black hole scaling relation \citep{Haehnelt,Peng} and to drive the size evolution \citep{Hopkins2009,Oser} of ETGs.
However, the picture is complicated by the many physical processes that are present during the evolution of galaxies, such as gas cooling, feedback from stars and AGN.
Moreover, an important fraction of the mass of ETGs is accounted for by dark matter \citep{Bertin,Franx,Gerhard,TreuKoop,Bar++11} whose nature is still unknown.
Understanding the interplay of baryonic and dark matter and how they act to produce the observed structural characteristics is essential to comprehend the evolution of ETGs, but is today a challenging task.

Theoretical models and simulations with a variety of physical ingredients have been set up to try to reproduce the observables of ETGs \citep{Gustafsson,Hopkins2010,Schaye,Duffy}.
Although simulations seem to be able to capture the general characteristics of ETGs \citep[e.g.][]{HopkinsHer,Ciotti2010}, quantitatively matching the entire set of observables proved to be difficult, often requiring an ad hoc tuning of the model parameters \citep{Nipoti,Hopkins2010}.
For example, \citet{Duffy} explored a variety of aspects of baryonic physics such as gas cooling, feedback from stars and AGN, finding that on the one hand the observed inner slopes of massive ETGs are reproduced if the feedback is weak, but on the other hand a strong feedback is needed to match the measured stellar masses.

Improving the quality of the observation of ETGs and introducing more constraints can help us to discriminate between the wealth of currently viable scenarios for their history.
Two characteristics of ETGs in particular are still not known with sufficient precision and leave room for significant improvement in their observational determination: the stellar mass and the density profile of the dark matter halo.
The stellar mass is degenerate with the stellar IMF with respect to constraints from the integrated light distribution and colors.
Breaking this degeneracy can help determining the star formation history and the content of the baryonic mass in ETGs.
The density profile of the dark matter distribution is sensitive to the physical processes that take place during the formation and evolution of ETGs.
Therefore, measuring the profile of dark matter halos in ETGs is a powerful means for testing the various theoretical models.

Part of the difficulty in comparing simulations with observations are, of course, due to the fact that dark matter, which accounts for a significant fraction of the mass of a typical galaxy, is not directly observable. 
Gravitational lensing is, in this aspect, a very powerful tool, being sensitive to the gravitational pull of matter independently on its interaction with light \citep[e.g.,][and references therein]{Tre10}.

Lensing surveys indeed played a crucial role in uncovering physical
characterics of early-type galaxies, such as their average density
profile and dark matter fraction \citep{Koop2006,Koo++09,PaperX,Bar++11} or the IMF of their stellar population
\citep{Grillo2009,TreuIMF,Auger2010L}.  Measurements of the inner
slope of a dark matter halo have so far been obtained for a few
cluster lenses \citep[e.g.][]{Sand08,Newman09}, for which constraints from
multiple lensed sources are available.  For typical early-type galaxy
strong lenses however, there are residual degeneracies between
anistropy, stellar mass to light ratio and inner slope of the dark
matter halo and therefore the constraints are weak
\citep{K+T03,TreuKoop}. For this reason, previous studies have adopted
theoretically motivated mass density profiles for the dark matter halo
\citep{TreuIMF,Auger2010L}, rather than free power laws.

Here we present a detailed study of an early-type galaxy at redshift
$z=0.222$.  The galaxy is the strong gravitational lens of the system
SDSSJ0946+1006, part of the SLACS sample \citep{Bolton2004}.  This ETG
is special in that it lenses two sources at different redshifts,
creating two nearly complete Einstein rings of different radii. For
this reason, the system is also referred to as the ``Jackpot''.  The
first lensed source is at redshift $z_{s1} = 0.609$, while there is no
spectroscopic measurement of the redshift of the second ring.  Thanks
to the presence of the two rings, this system provides more
information than typical gravitational lenses, despite the lack of the second source redshift. 
A first study of SDSSJ0946+1006 was carried out by \citet[Paper VI]{Gavazzi}.  
An independent lensing analysis of this system was performed by
\citet{Vegetti}, which led to the discovery of a small satellite with no visible counterpart.
Here we include new high-quality photometry obtained
with the Hubble Space Telescope (hereafter HST) and new deep and
spatially resolved spectroscopy obtained at the Keck Telescope.  The
goal of our study is to separate the contribution of dark and stellar
matter to the total mass of the lens, making as few assumptions as
possible about the density profile of the dark matter halo.  This task
is achieved by combining lensing and dynamics information.  Unlike
typical early-type galaxy lenses, the wealth of information provided
by this system allows us to determine both the mass of the stellar
bulge and the inner slope of the dark matter halo.  Thanks to the
multi-band HST photometry we are able to obtain a photometric redshift
of the outer ring, that is necessary for improving the constraints
from the lensing data, and to infer stellar masses from stellar
population synthesis (SPS) fitting.  The comparison between this
measurement of the stellar mass and the one obtained through lensing
and dynamics allows us to constrain the IMF of the stars in the lens.
This is the most robust measurement of the inner slope of the dark
matter halo and IMF of an isolated massive ETG.

The structure of this paper is the following.  In Sections 2 and 3 we
describe the new photometric and spectroscopic data, respectively.
Our measurement of the photometric redshift of the outer ring is
presented in Section 4.  In Section 5 we describe measurements of the
stellar mass of the lens from stellar population synthesis fitting.
Section 6 describes a lensing and dynamics model assuming a power-law
density profile for the total density profile of the lens.  In Section
7 we present the bulge-halo decomposition of the lens.  We discuss our
results in Section 8 and summarize in Section 9.

Throughout the paper we assume the following values for the cosmological parameters: $H_0 = 70\mbox{ km s}^{-1}\mbox{Mpc}^{-1}$, $\Omega_M = 0.3$, $\Omega_\Lambda = 0.7$. Magnitudes are expressed in the AB system, images are North-up and position angles are in degrees East of North.
In showing our results we display posterior PDFs in multiple projections wherever possible, but when
giving a point estimate of an inferred parameter we quote the position
of the peak of its one-dimensional marginalised distribution, with
uncertainties defined by the 68\% credible region.

\section{Multicolor HST photometry}

\begin{figure}
\includegraphics[width = \columnwidth]{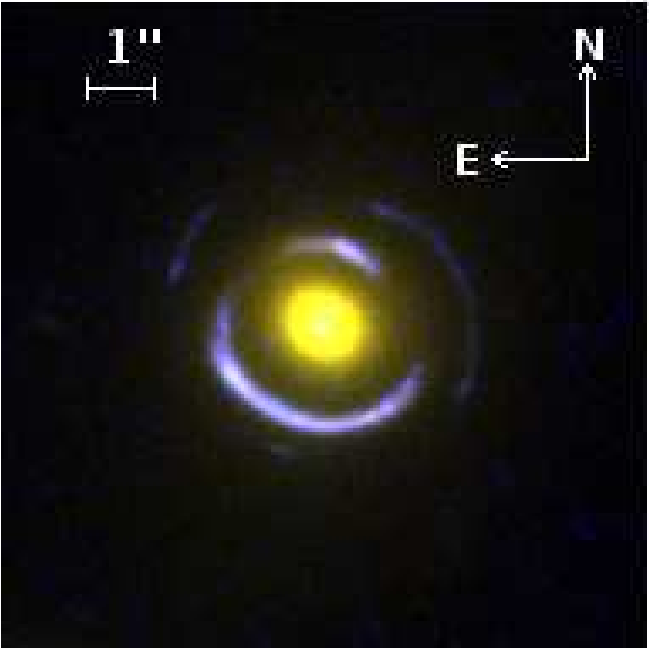}
\caption{Gravitational lens system SDSSJ0946+1006 in a combination of F814W, F606W and F336W HST images.}
\label{fig:Jackpot}
\end{figure}

We present HST images of the lens system SDSSJ0946+1006 in five
different bands.  In Paper VI we reported results based on an ACS
F814W image only. Images in WFPC2 F606W and WFC3-IR F160W (Cycle 16,
Program 11202, PI Koopmans) were available for Paper IX \citep{PaperIX}.  In addition
to those data, we now have WFC3 images in F438W and F336W bands (Cycle
17, Program 11701, PI Treu).  Table \ref{imgtable} summarizes the
observations.  This section describes the data reduction process
(\S~\ref{Reduction}) and the photometric properties we derived for the
lens galaxy (\S~\ref{LensPhoto}).  For conciseness, we sometimes refer
to the F160W, F814W, F606W, F438W, F336W bands as H, I, V, B, U
respectively. A color composite image of the lens system is shown in
Figure~\ref{fig:Jackpot}

\begin{table}
\begin{center}
\caption{Summary of the HST observations.\label{imgtable}}
\begin{tabular}{ccccc}
\tableline\tableline
Instrument & Filter & Exp. time & $N_{exp}$ & Date \\
\tableline
WFC3 IR & F160W & 2397 s& 4 & 09/12/2009\\
ACS & F814W & 2096 s & 4 & 3/11/2006 \\
WFPC2 & F606W & 4400 s & 4 & 18/12/2009 \\
WFC3 UVIS & F438W & 2520 s & 4 & 20/03/2010\\
WFC3 UVIS & F336W & 5772 s & 4 & 20/03/2010\\
\tableline
\end{tabular}
\end{center}
\end{table}

\subsection{Data reduction}\label{Reduction}

The data are treated with the standard HST reduction pipeline. For each image, frames are coadded and resampled in a uniform pixel scale using the software {\sc multidrizzle} \citep{Drizzle}. Pixel sizes are $0.10''$ for the F160W image, $0.050''$ for the F814W and F606W images, and $0.0396''$ for the F438W and F336W images.
The images are then brought to the same orientation and $0.050''$ pixel scale by using the software {\sc swarp} \citep{Swarp}.
The PSF of each image is estimated from stars in the field.

\subsection{Lens galaxy properties}\label{LensPhoto}

\begin{figure*}[!]
  \begin{center}
    \begin{tabular}{ccc}
      \includegraphics[width=0.31\textwidth]{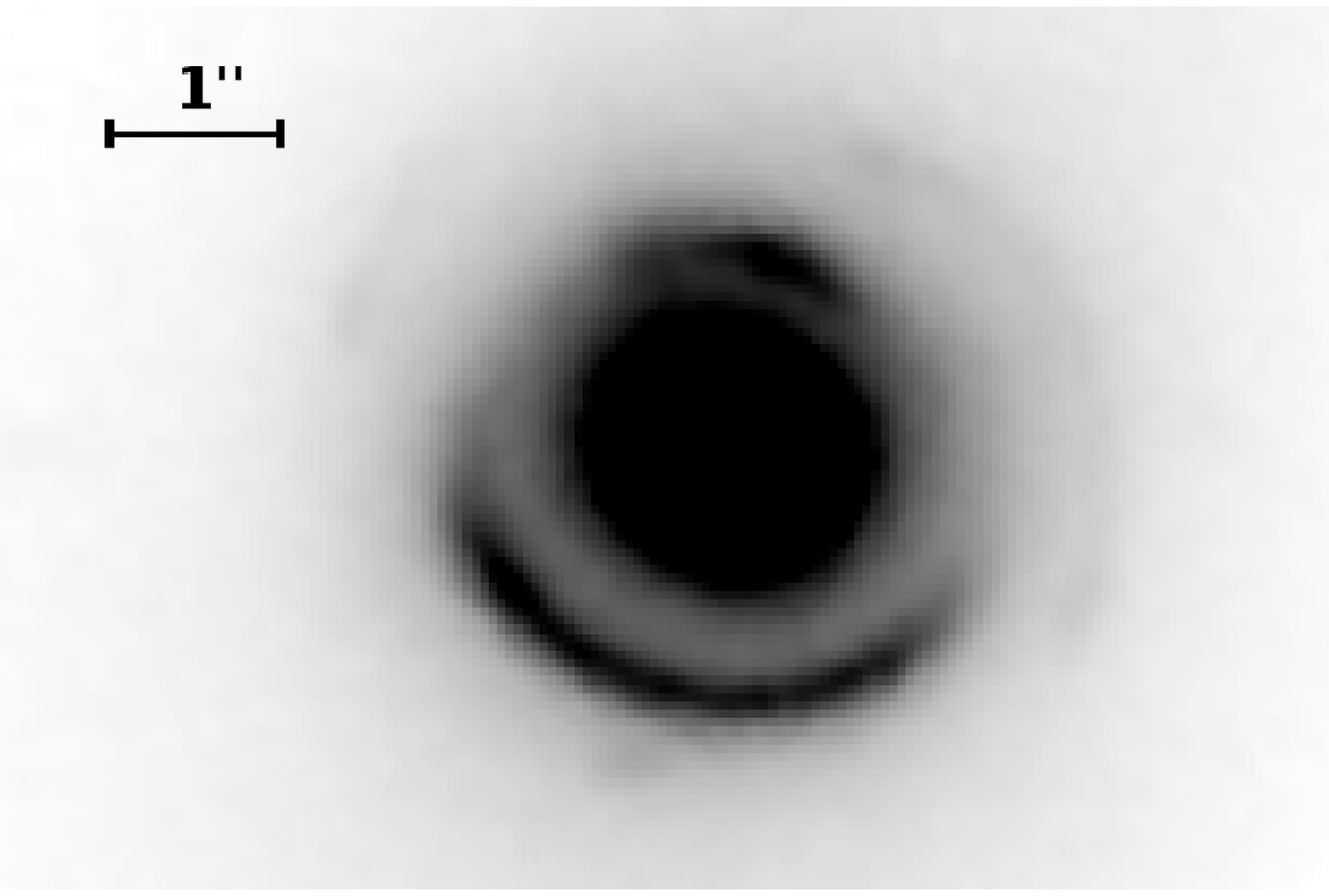} &
      \includegraphics[width=0.31\textwidth]{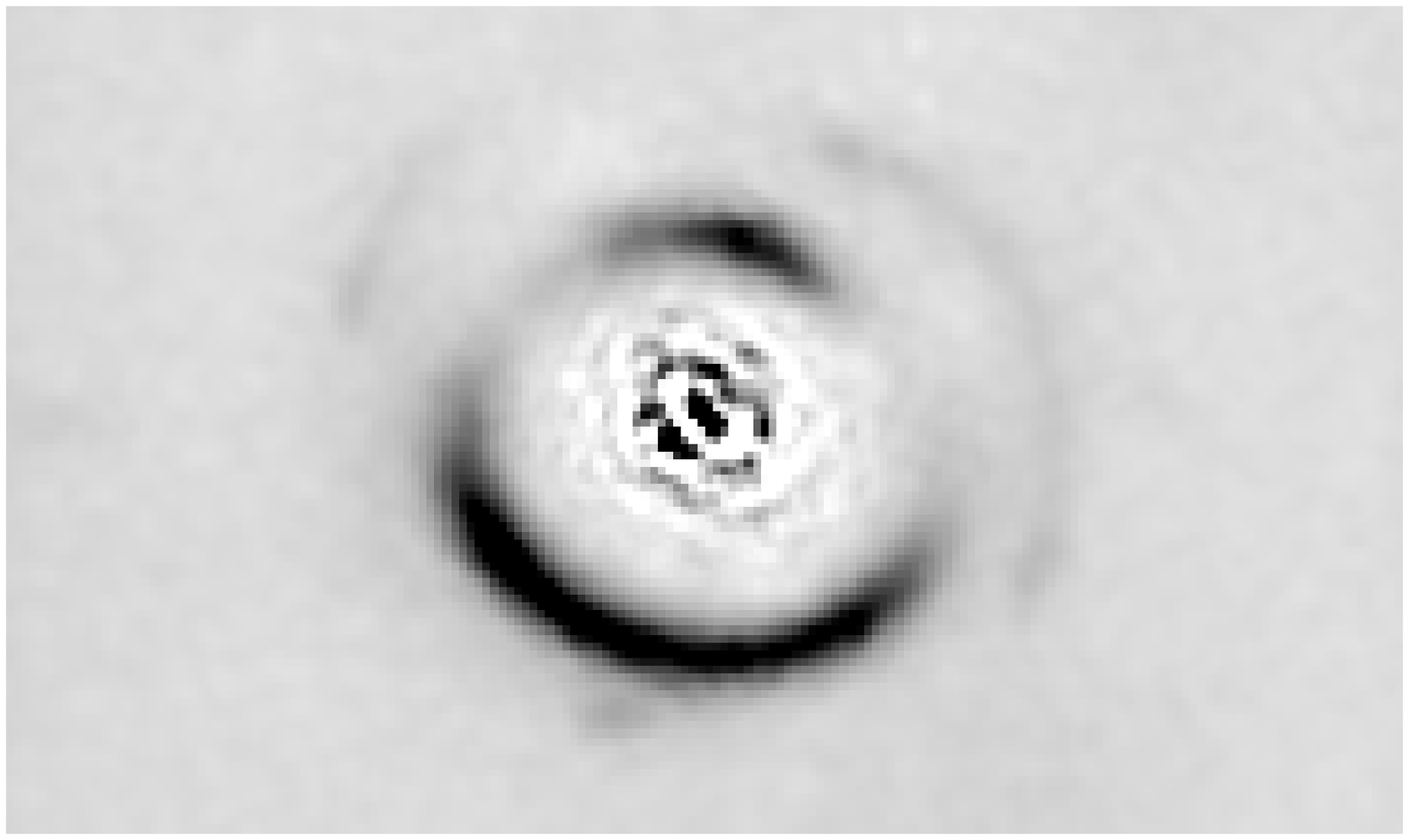} &
      \includegraphics[width=0.31\textwidth]{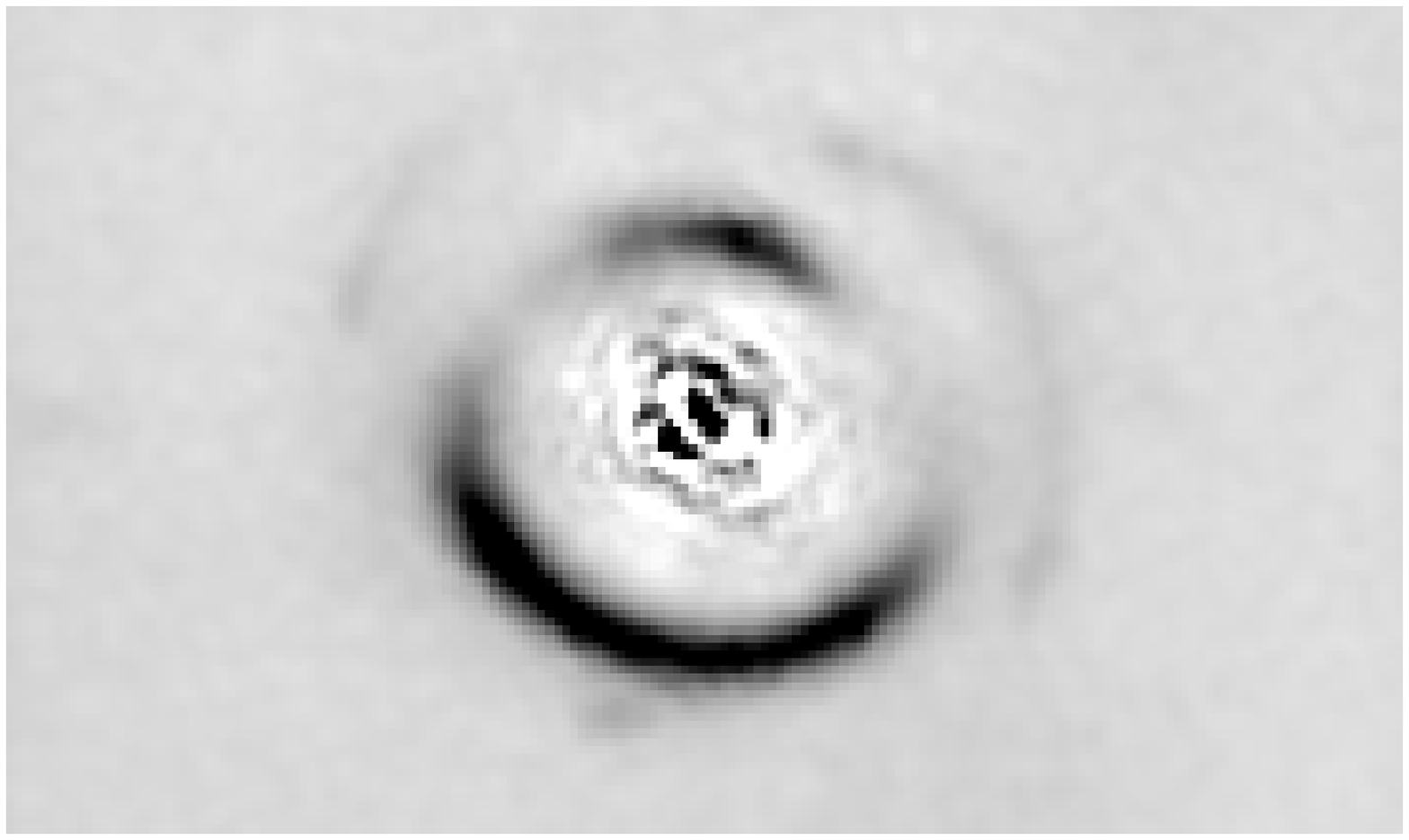} \\
      \includegraphics[width=0.31\textwidth]{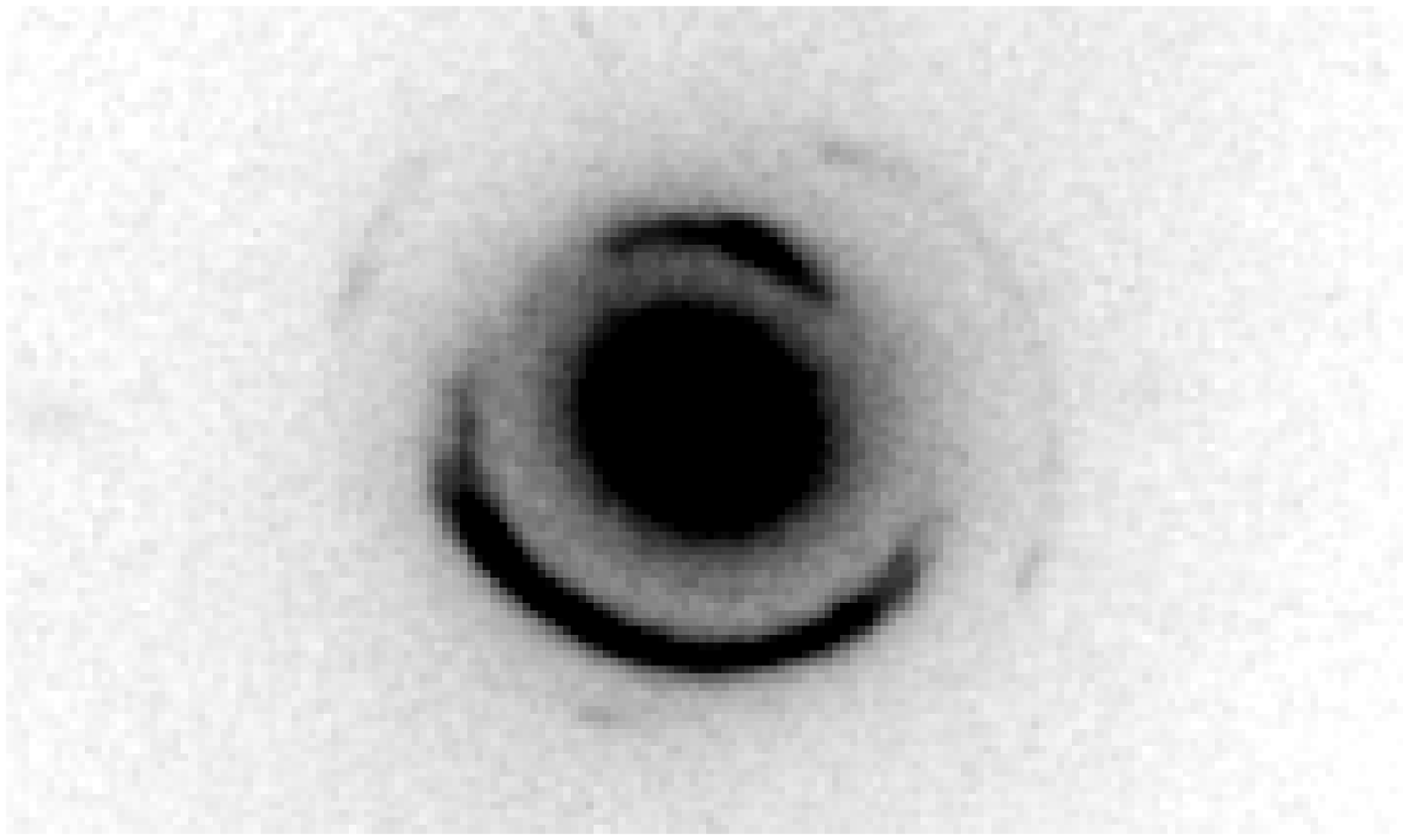} &
      \includegraphics[width=0.31\textwidth]{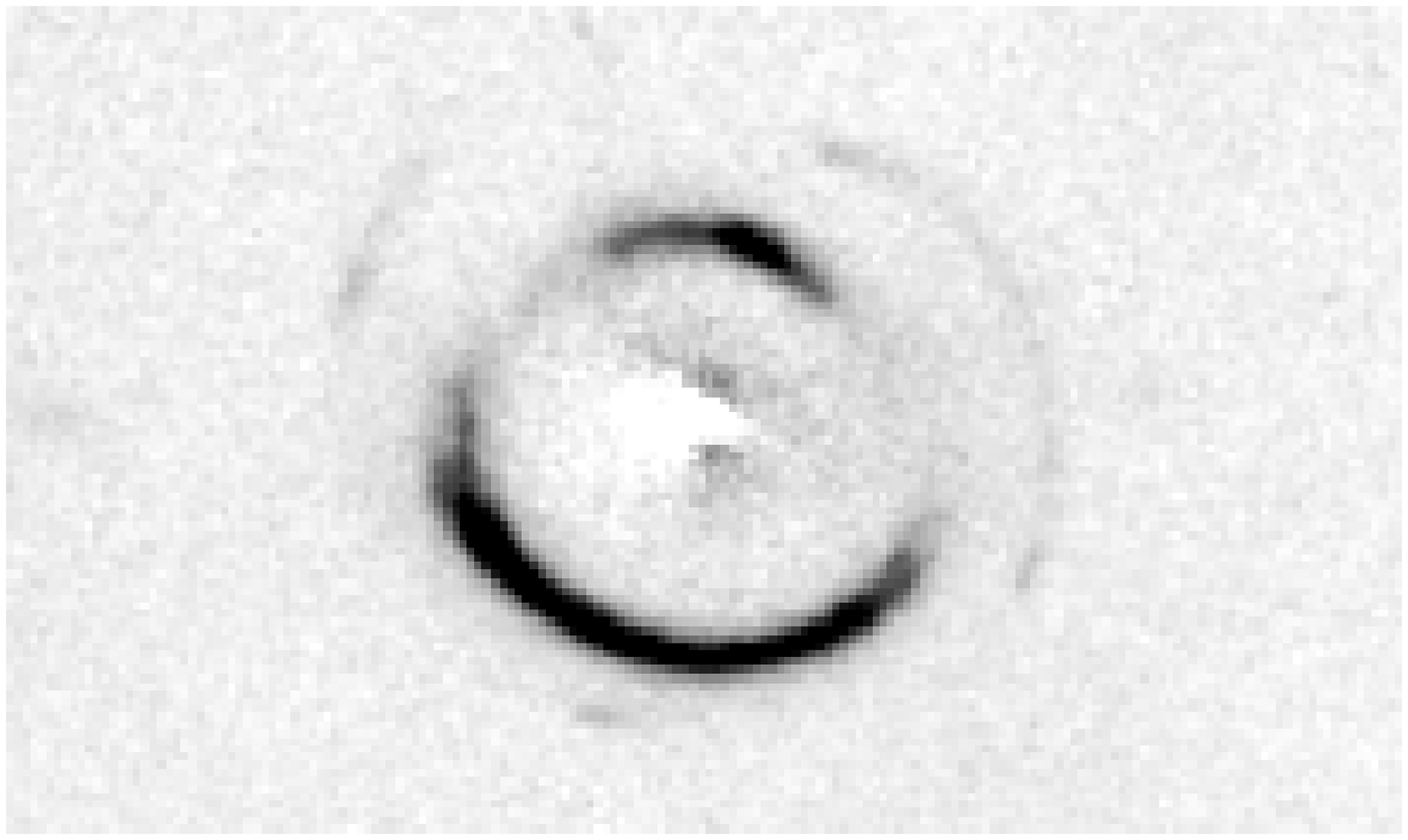} &
      \includegraphics[width=0.31\textwidth]{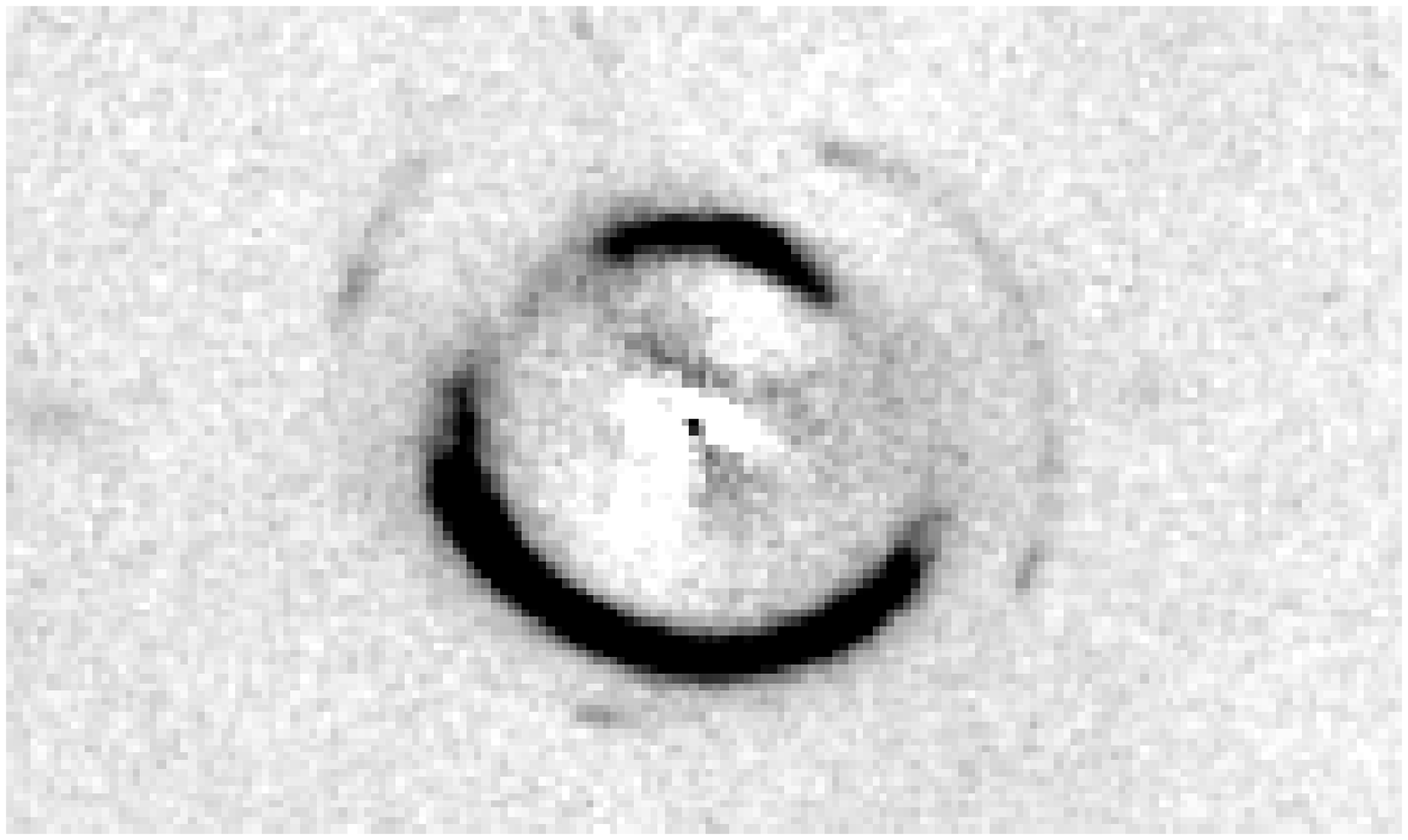} \\
      \includegraphics[width=0.31\textwidth]{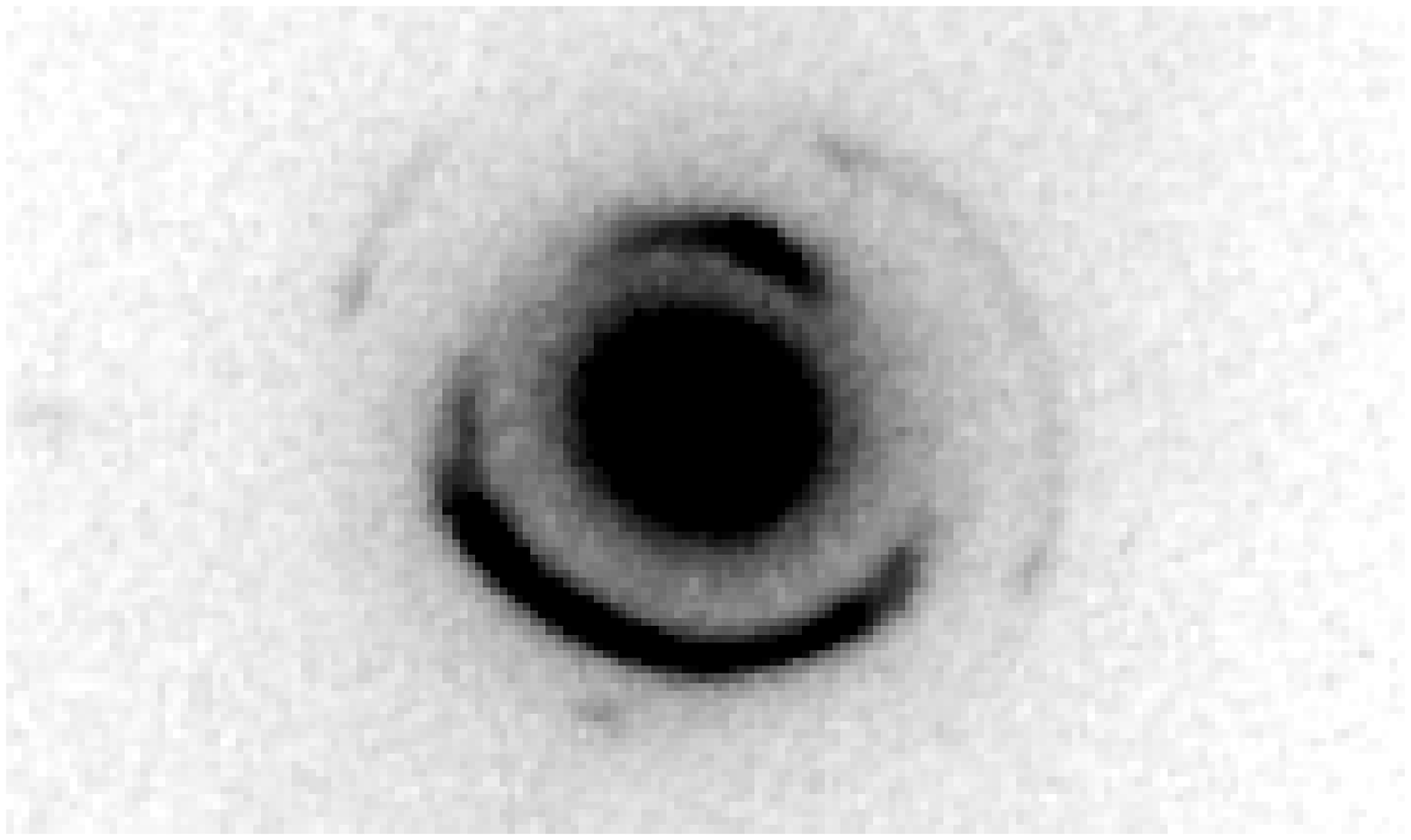} &
      \includegraphics[width=0.31\textwidth]{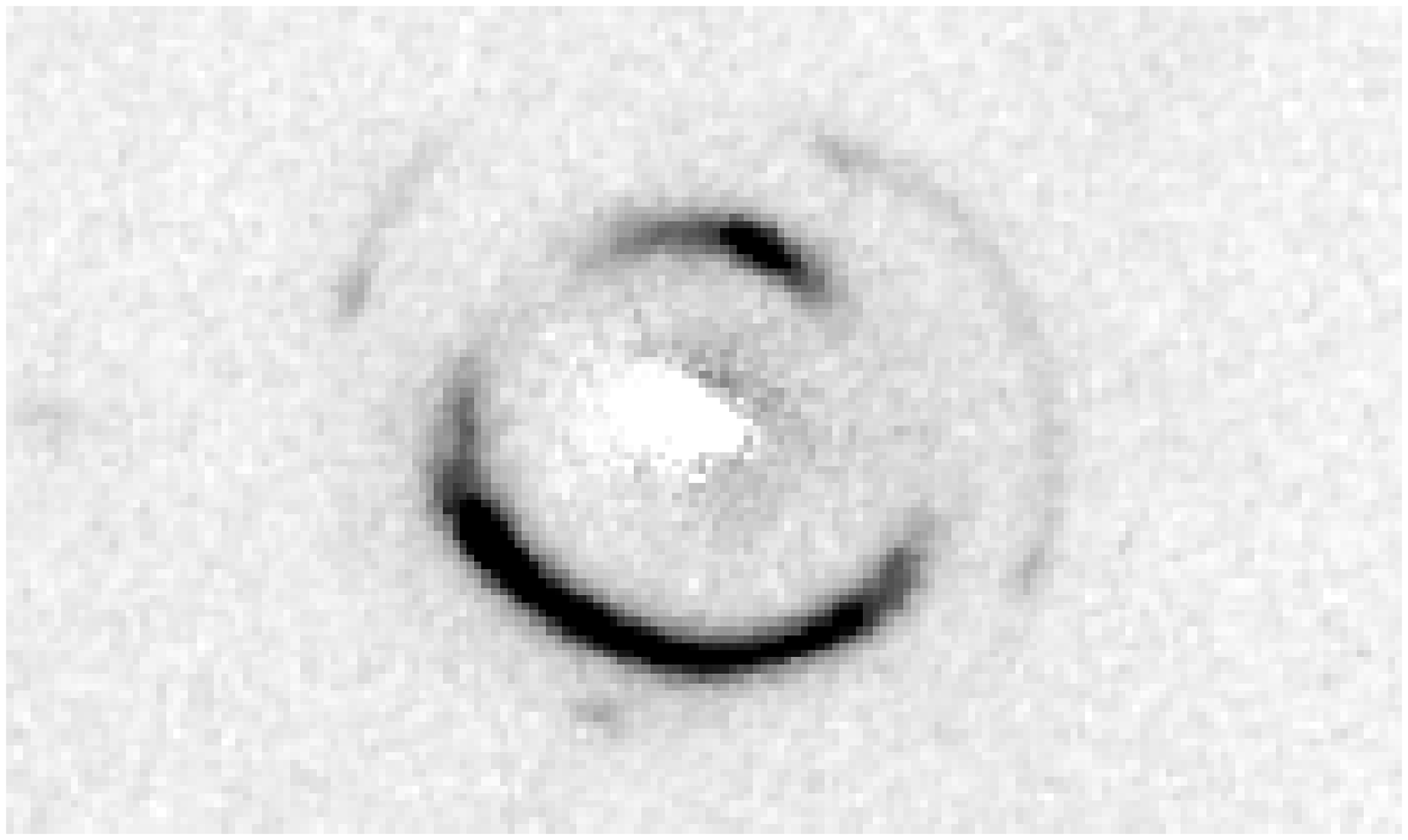} &
      \includegraphics[width=0.31\textwidth]{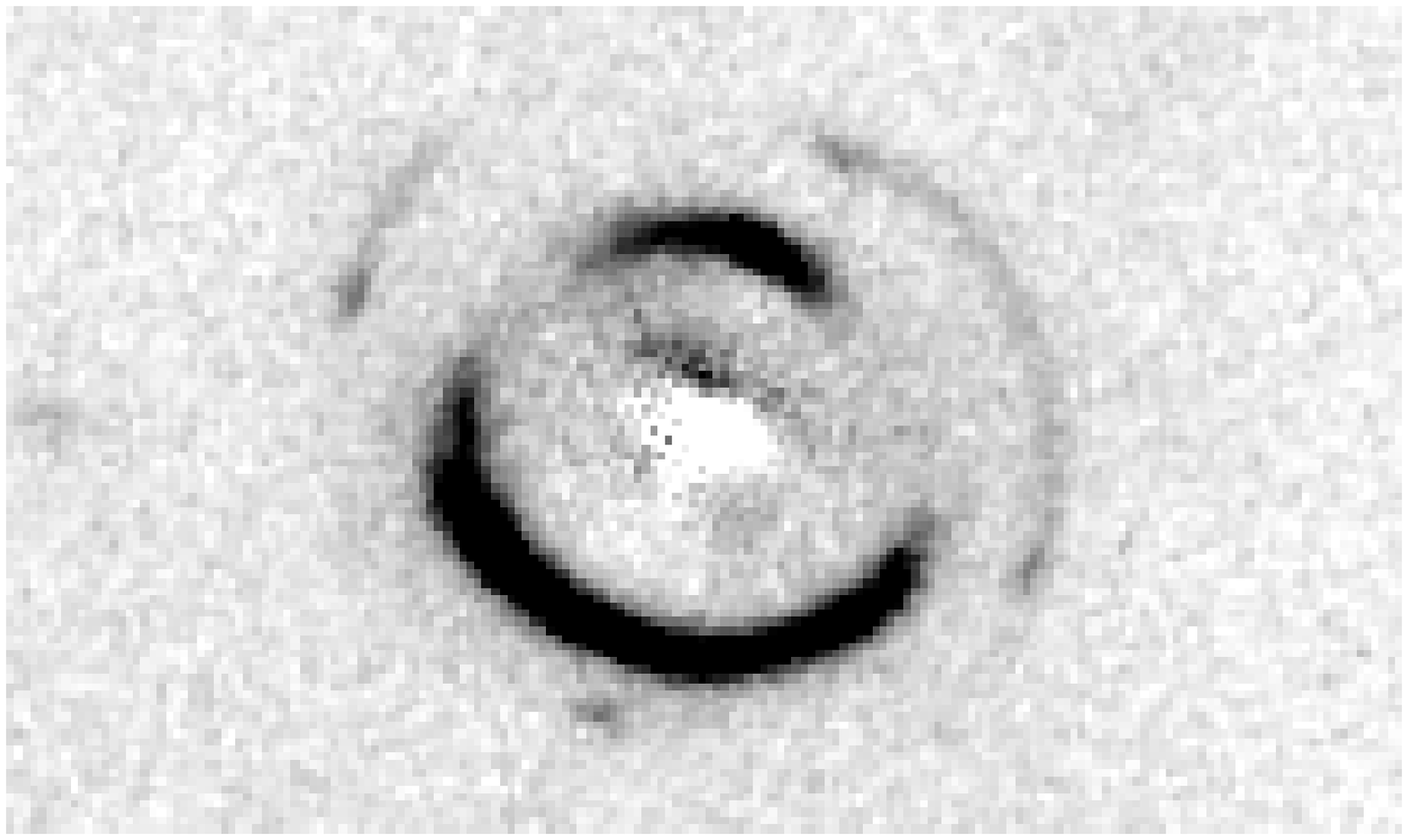} \\
      \includegraphics[width=0.31\textwidth]{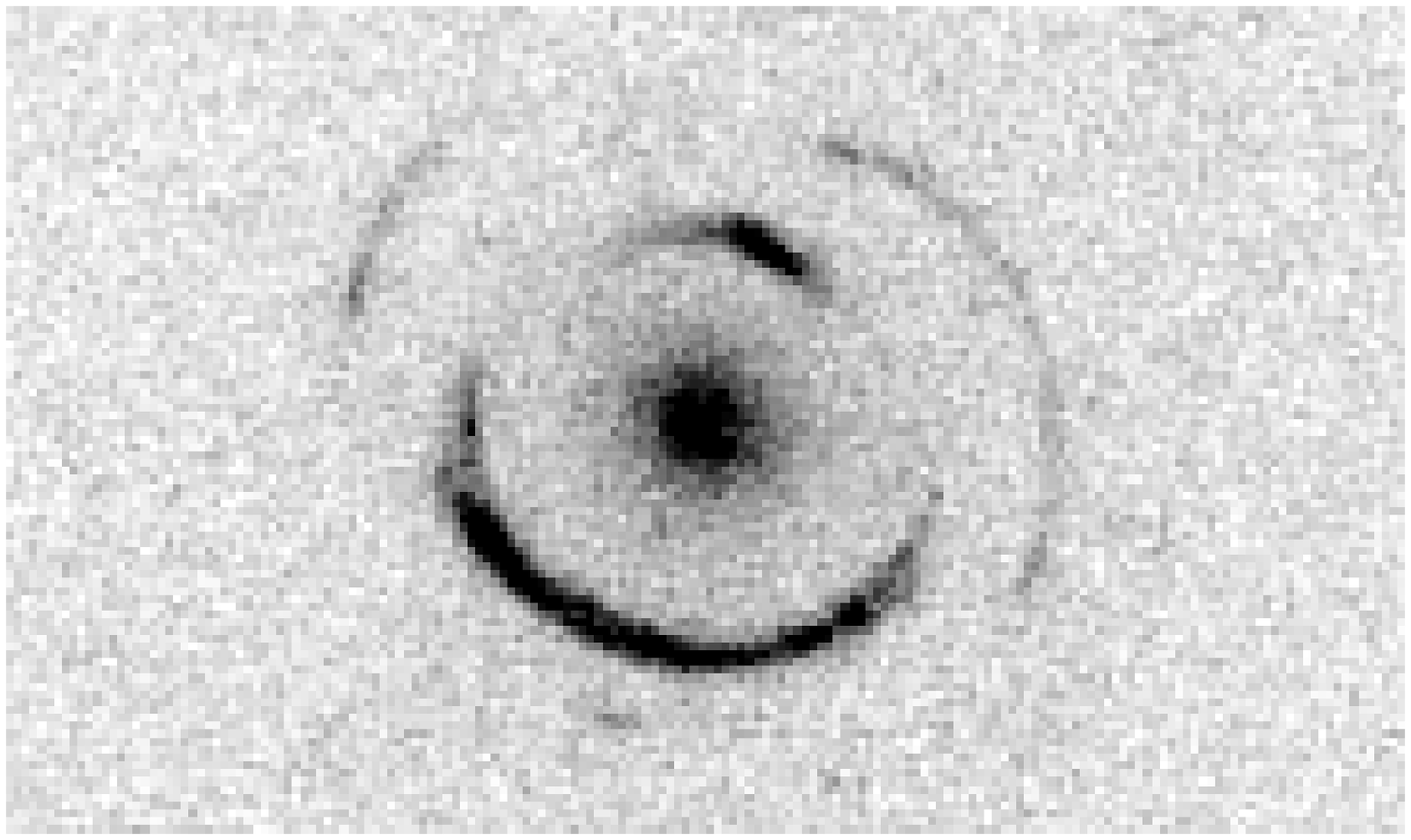} &
      \includegraphics[width=0.31\textwidth]{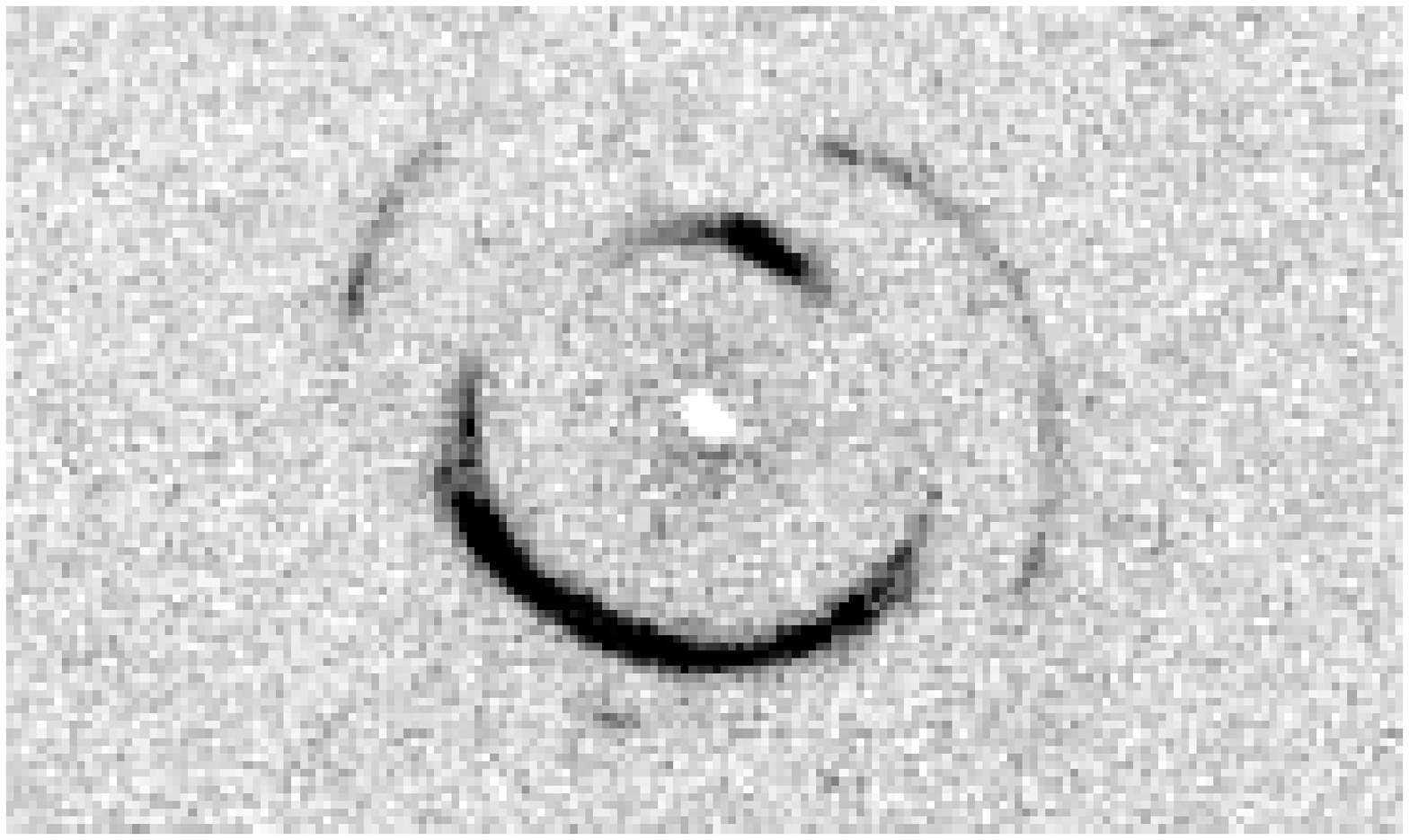} &
      \includegraphics[width=0.31\textwidth]{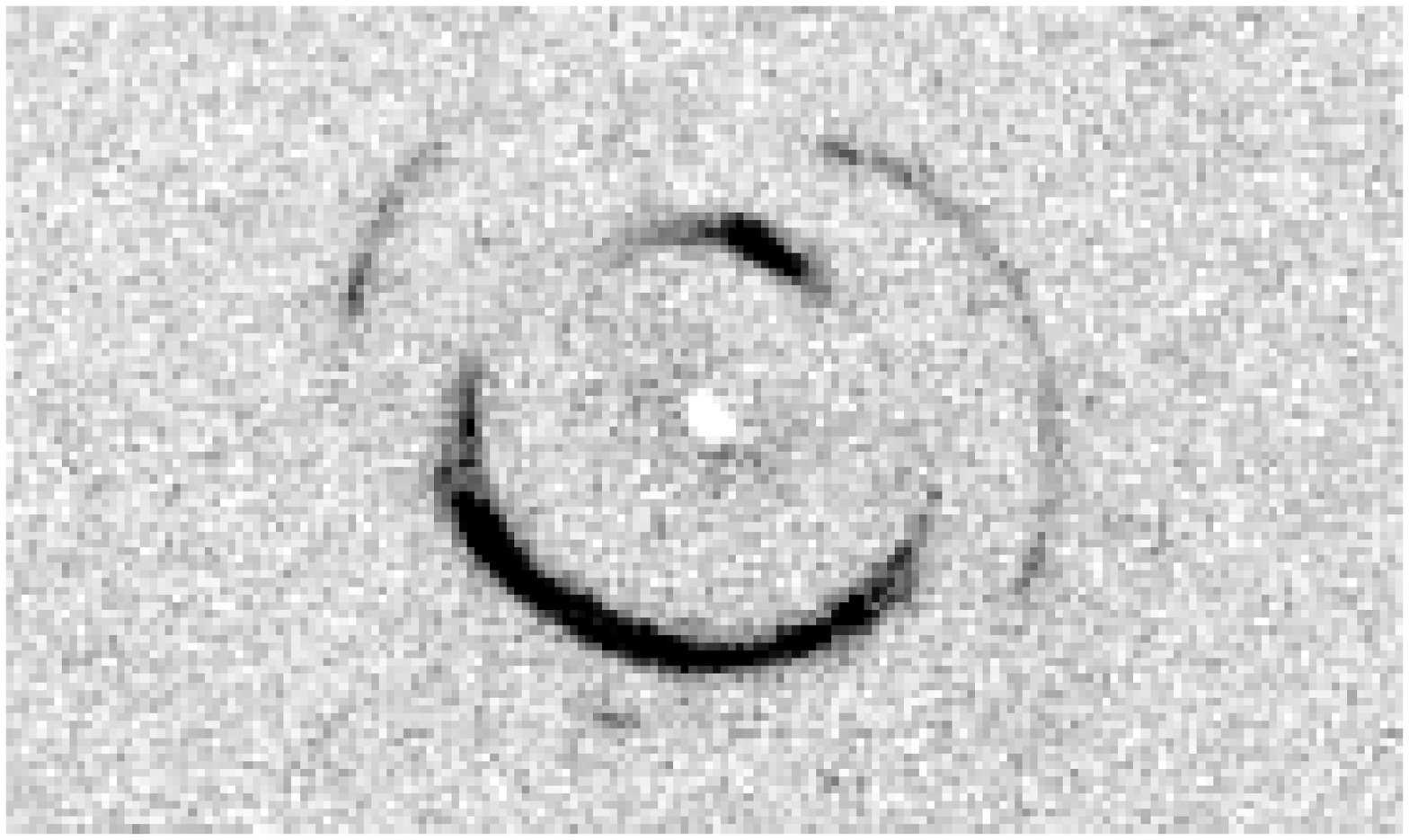} \\
      \includegraphics[width=0.31\textwidth]{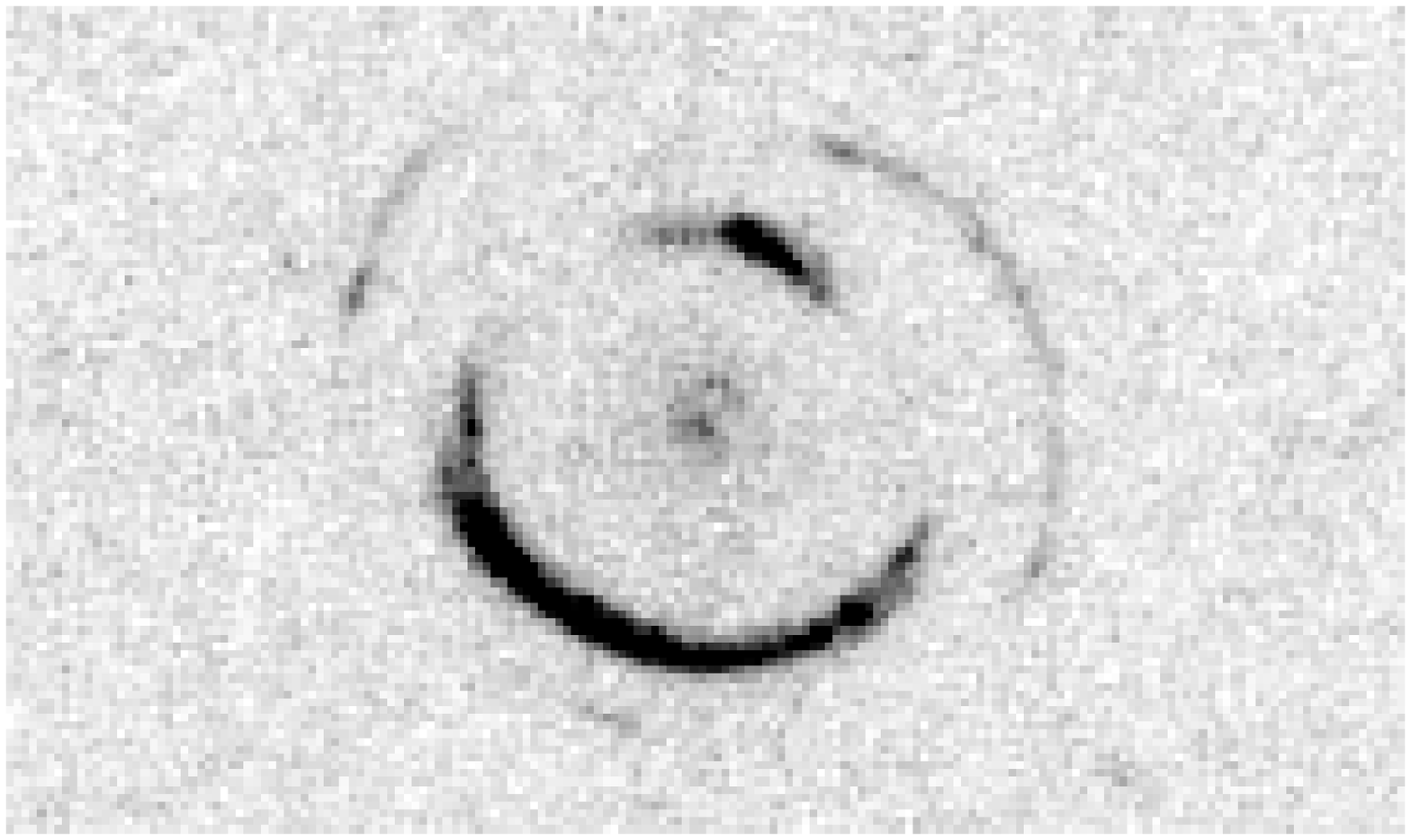} &
      \includegraphics[width=0.31\textwidth]{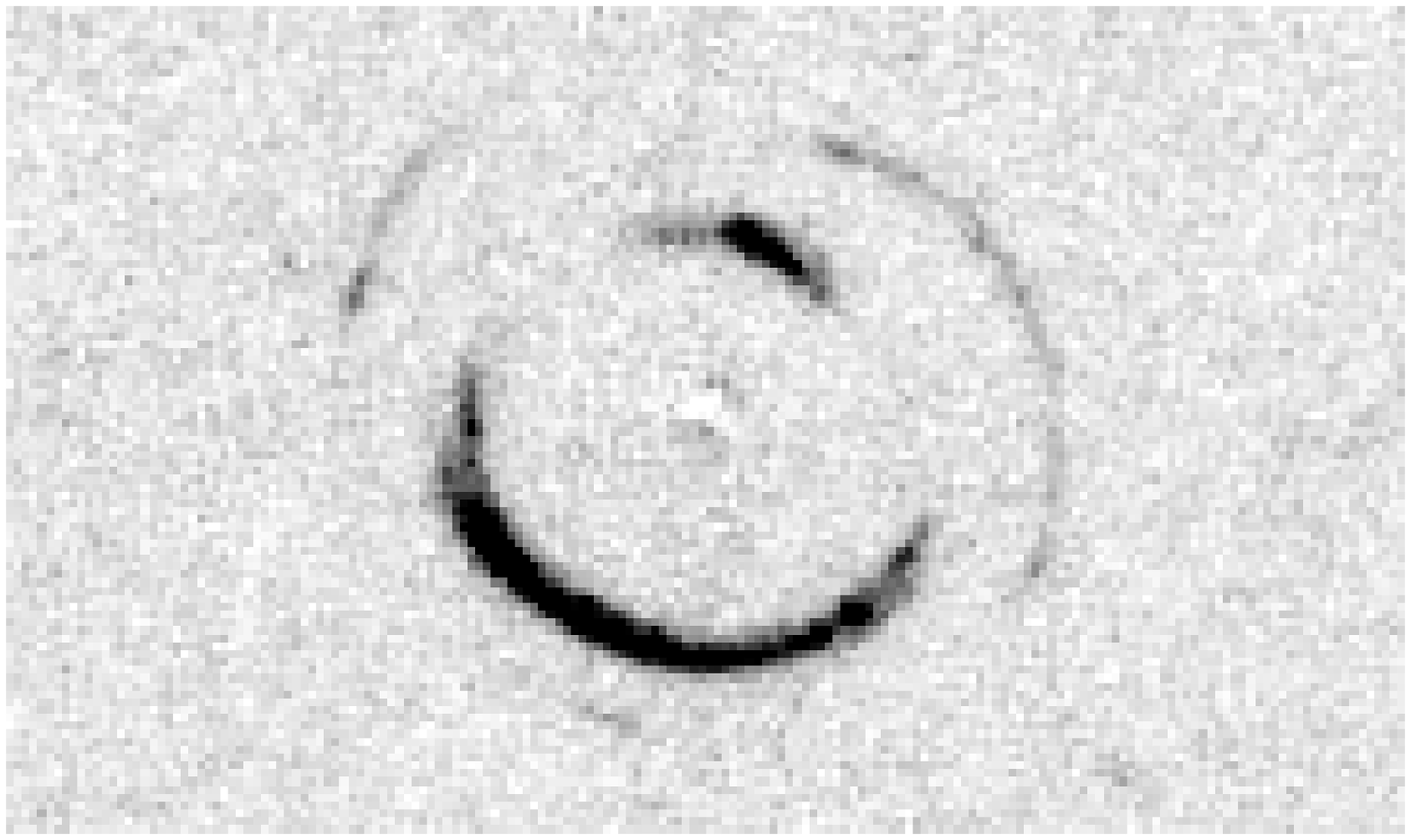} &
      \includegraphics[width=0.31\textwidth]{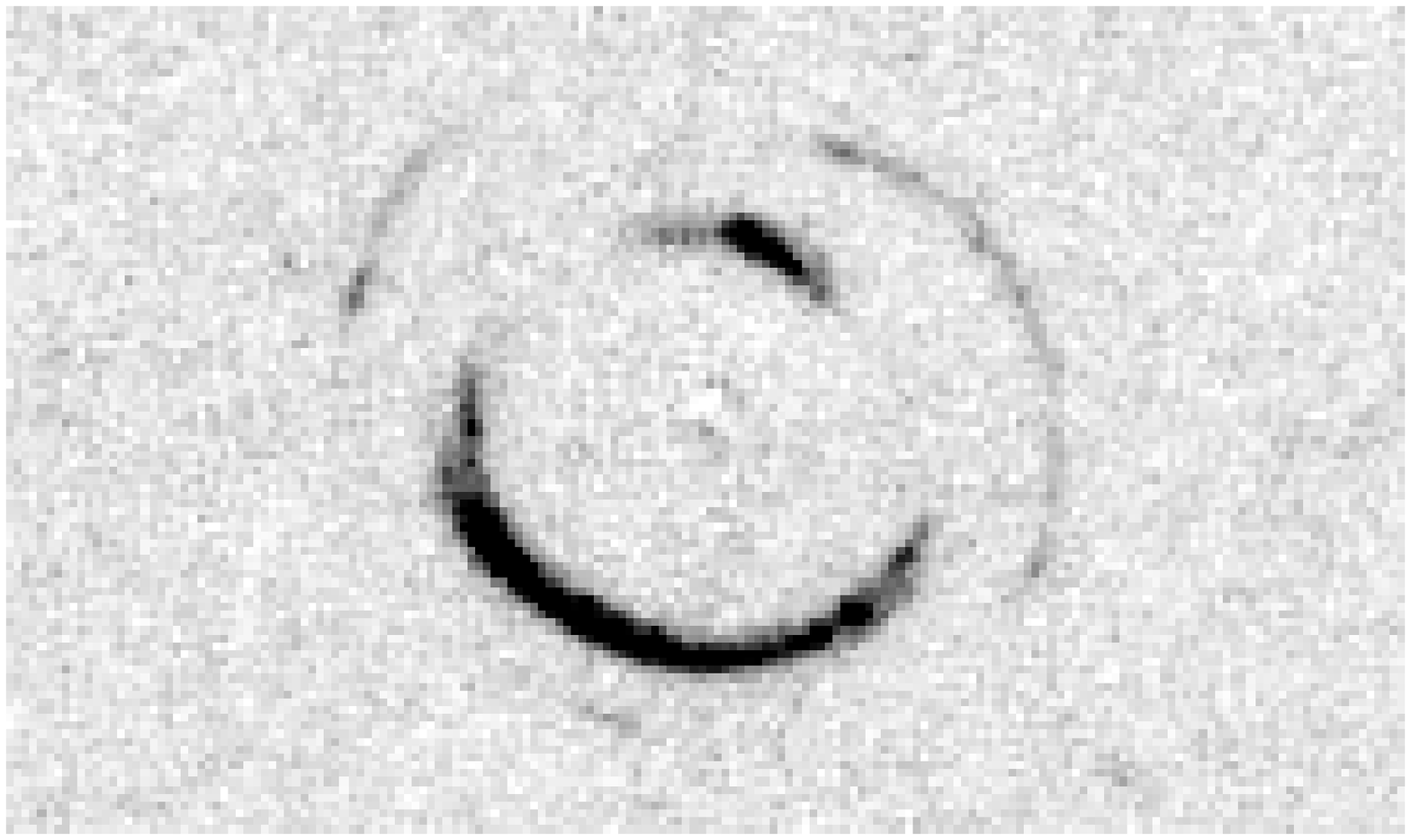} \\
    \end{tabular}
    \caption{From top to bottom: HST F160W, F814W, F606W, F438W and F336W images of the lens system SDSSJ0946+1006 before (left column) and after (middle and right column) light subtraction. {\em Middle column:} light distribution modeled as a double S\'{e}rsic profile, with parameters given in Table \ref{TableFits}.
{\em Right column:} light distribution modeled as a double tPIEMD profile, with parameters given in Table \ref{twotNIEfit}.}
    \label{imgs1}
  \end{center}
\end{figure*}

The brightness distribution of the main lens galaxy is first obtained
by fitting S\'{e}rsic profiles to the data. This task is achieved with
the software {\sc spasmoid}, developed by M. W. Auger and described by
\citet{Bennert}. {\sc Spasmoid} fits the data in all the bands
simultaneously with a unique model, determining total magnitude and
colors of the galaxy at once.  
By using a single S\'{e}rsic component we find a best-fit profile described by a S\'{e}rsic index $n=6.0$, axis ratio $q = 0.95$ and effective radius $r_{eff} = 2.93''$.
However, the residuals left by this single-component fit are rather large.
Consequently, we add a second
component, allowing for the position angle of the major axes of the
two profiles to be different but imposing a common centroid.  In the fitting process, the light from
the rings is masked out manually.  This procedure gives robust
estimates of the colors of the lens, rather independent from the model
adopted to describe the data.  Color information will be used in
Section 5 to constrain the stellar population.  In Fig. \ref{imgs1} we
show the images of the system in the five bands, before and after
subtracting the main lens. 
Residuals are on the order of a few percent in the F814W band image.
Table \ref{TableFits} reports the best-fit structural parameters of the
model, while the best-fit colors are given in Table \ref{colortable}.
It is worth pointing out that the major axes of the two components are
almost perpendicular, and that the mean surface brightness within the
effective radius of component 1 is a factor $\sim30$ larger than that
of component 2.
The measured magnitude in the F814W band is consistent with the value reported by \citet{Gavazzi} for the same object.

In order to both explore model-dependent systematic errors and obtain a computationally more tractable description of the light profile for our lensing analysis, we also model the lens light with the following surface brightness distribution:
\begin{equation}\label{tNIEsb}
I(x,y) = I_cr_c\left[\frac{1}{\sqrt{r_c^2+R^2}} - \frac{1}{\sqrt{r_t^2+R^2}}\right],
\end{equation}
where $R^2 \equiv x^2/q + qy^2$. 
This profile corresponds to a truncated pseudoisothermal elliptical mass distribution (tPIEMD) in 3d, with $r_c$ and $r_t$ corresponding to the core radius and truncation radius respectively.
Note that the number of parameters of the model is the same as that of a S\'{e}rsic profile.
Two components are used, as in the S\'{e}rsic case.
The best-fit parameters are reported in Table \ref{twotNIEfit}.
Both the double-S\'{e}rsic and the double-tPIEMD profiles fit well the photometry of the lens, with residuals within the outer ring on the order of a few percent in the F814W band (see Figure \ref{imgs1}).

The inferred total magnitude in the two models is different, but this is due to the different behavior at large radii, where there are no data.
In fact, the magnitude within the inner ring is the same for the two models to within 0.01 mags and the inferred colors are consistent within the errors with those reported in Table \ref{colortable}.

\begin{deluxetable*}{ccccccc}
\tablewidth{0pt}
\tablecaption{Lens light distribution: double S\'{e}rsic model\label{TableFits}}
\tablehead{
\colhead{Component} & \colhead{$m_{\rm{F814W}}$} & \colhead{$r_{\rm{eff}}$} & \colhead{$n$} &
\colhead{$q$} & \colhead{PA} & \colhead{$\left<\mathrm{SB}\right>_{\rm{e,F814W}}$}\\
& (mag) & (arcsec) & & & (degrees) & (mag arcsec$^{-2}$)}
\startdata
1 & $18.38\pm0.20$ & $0.50\pm0.10$ & $2.34\pm0.50$ & $0.79\pm0.10$ & $63.0\pm1.0$ & $18.87\pm0.10$\\
2 & $17.44\pm0.10$ & $4.46\pm0.50$ & $1.60\pm0.50$ & $0.64\pm0.10$ & $-23.4\pm1.0$ & $22.68 \pm 0.20$\\
\enddata
\tablenotetext{}{Best fit parameters for the double-S\'{e}rsic model surface brightness profile of the main lens: magnitude in the F814W band, effective radius, S\'{e}rsic index ($n$), axis ratio ($q$), position angle of the major axis (East of North), effective surface brightness. Each line refers to one of the S\'{e}rsic components of the model.
The errors represent the typical range of values for the parameters allowed by the model. 
These errors are correlated: for example, an increase in the value of the S\'{e}rsic index $n$ results in a change of the effective radius to fit the observed slope in surface brightness.}
\end{deluxetable*}

\begin{deluxetable*}{ccccccc}
\tablewidth{0pt}
\tablecaption{Lens light distribution: double tPIEMD model\label{twotNIEfit}}
\tablehead{
\colhead{Component} & \colhead{$m_{\rm{F814W}}$} & \colhead{$r_c$} & \colhead{$r_t$} &
\colhead{$q$} & \colhead{PA} & \colhead{$\left<\mathrm{SB}\right>_{\rm{e,F814W}}$}\\
& (mag) & (arcsec) & (arcsec) & & (degrees) & (mag arcsec$^{-2}$)}
\startdata
1 & $18.75\pm0.20$ & $0.066\pm0.010$ & $0.50\pm0.05$ & $0.66\pm0.10$ & $63.0\pm1.0$ & $19.24\pm0.10$\\
2 & $17.15\pm0.10$ & $0.082\pm0.010$ & $6.05\pm0.10$ & $0.71\pm0.10$ & $-24.3\pm1.0$ & $22.48 \pm 0.20$\\
\enddata
\tablenotetext{}{Best fit parameters for the double-tPIEMD model surface brightness profile of the main lens: magnitude in the F814W band, core radius ($r_c$), truncation radius ($r_t$), axis ratio ($q$), position angle of the major axis (East of North), effective surface brightness.}
\end{deluxetable*}

\begin{table}
\begin{center}
\caption{Colors of the lens galaxy.\label{colortable}}
\begin{tabular}{cccc}
\tableline\tableline
Color & Component 1 & Component 2 & Global \\
\tableline
I - H & $1.16\pm0.05$ & $0.86\pm0.05$ & $0.96\pm0.05$\\
V - I &  $0.81\pm0.05$ & $0.96\pm0.05$ & $0.91\pm0.05$\\
B - V & $2.36\pm0.20$ & $1.52\pm0.05$ & $1.73\pm0.05$\\
U - B & $2.30\pm0.30$ & $1.32\pm0.10$ & $1.44\pm0.10$\\
\end{tabular}
\end{center}
\end{table}

\begin{figure}
\includegraphics[width = \columnwidth]{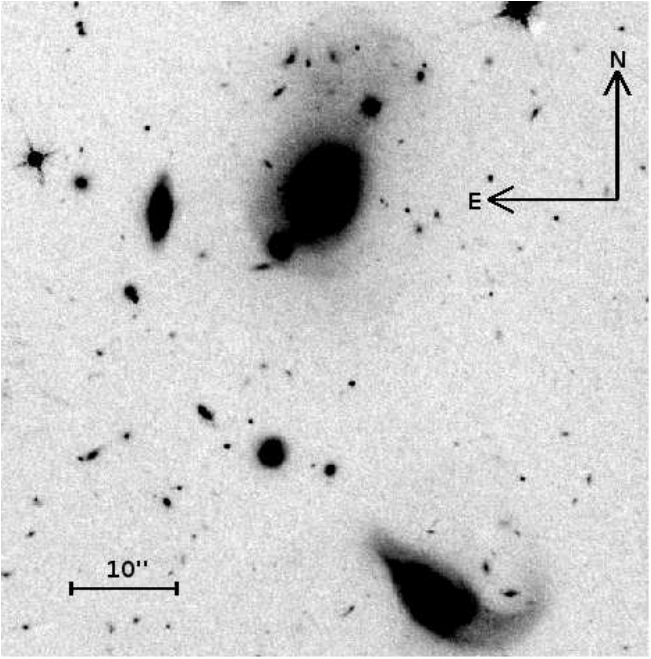}
\caption{WFC3-IR F160W image of the lens system and its surroundings. Note the irregular shape of the faint stellar component at the outskirts of the lens galaxy (top of the image).
At the bottom, a neighbor also shows signs of tidal disruption. Both these features may be the result of a close encounter between the two objects.
}
\label{IRtidal}
\end{figure}

The infrared F160W data reveal distorsions in the shape of the light
distribution at large radii (see Figure \ref{IRtidal}), a possible
signature of tidal interactions.  
As previously noted by \citet{Gavazzi}, a galaxy in the neighborhood of the lens also shows signs of a tidal interaction (see Figure \ref{IRtidal}). It is possible that the two galaxies are undergoing a merger.
This deviation from a regular light
profile is located far from the probed by our lensing and
dynamics measurements and is therefore not a concern for the accuracy
of our models.
The central part appears smooth to the few percent level and it is unlikely that the ongoing interaction would have an effect on its structure, given its deep potential well.
However, as we will discuss in \S~\ref{ssec:form} this
feature provides an interesting clue to the formation mechanism of
this galaxy.

Another interesting feature is revealed by the image in the F336W (U) band, as there seems to be some structure in the center of the lens (see Figure \ref{imgs1}).
The fact that this feature is clearly visible only in the U band, where the lens is fainter, may suggest that it is in fact a bluer object distinct from the central galaxy, or blue emission from an active nucleus. Alternatively, the observed detail could be the result of the presence of a dust lane that separates the light of the lens into two components at shorter wavelengths. In principle it could also be an additional image of the lensed sources.

One way to discriminate between a blue object or a dust lane is to
study the position of the centroid of the lens in the different bands.
A blue object would shift the centroid towards itself at bluer
wavelengths, while a dust lane would remove blue light, causing the
apparent centroid to move away from it.  When fitting for the centroid
of the lens, this latter case is observed: the centroid moves by about
one pixel towards the S-E in the F336W and F438W bands with respect to
the F814W band. This is a significant effect given the subpixel
accuracy of centroniding, and it suggests that dust is most likely the
cause of the observed feature in the F336W band.  
A more detailed discussion of the dust issue is given in Appendix A.

\section{Photometric redshift of the outer ring}\label{RingsPhoto}

\subsection{Colors of the ring}

One of the main goals of this study is to constrain better the mass distribution in the lens galaxy by obtaining a photo-z of the outer ring.
This task requires a measurement of the colors of the ring.
A color map of the outer ring is obtained as follows.
For each pair of neighboring bands, $\lambda_1$, $\lambda_2$, we align the corresponding images and then convolve each image with the PSF of the neighboring band. In this way we obtain pairs of images with the same effective PSF, necessary to get an unbiased estimate of the color for each pixel.
Global colors are then measured in the following way. For a given pair of bands, we select individual pixels with flux larger than the background by more than two sigma in both of the bands considered. 
We make the assumption that the source has spatially uniform colors and estimate them statistically by taking a weighted mean of the individual pixel colors.
The measured values of the colors, corrected for galactic extinction, are reported in Table \ref{TableRingColors}.

\begin{table}
\begin{center}
\caption{AB colors of the outer ring\label{TableRingColors}}
\begin{tabular}{cc}
\tableline\tableline
I - H & $0.61\pm0.10$ \\
V - I & $0.21\pm0.10$ \\
B - V & $0.15\pm0.10$ \\
U - B & $0.53\pm0.10$ \\
\end{tabular}
\end{center}
\end{table}

\subsection{Measuring the photo-z}\label{BPZ}

To estimate the photometric redshift of the outer ring we make use of
the software BPZ \citep[Bayesian Photo-z;][]{Benitez}.  Photo-z
analysis consists of fitting synthetic SEDs to the observed
colors. BPZ works in a Bayesian framework that allows us to combine
the inference with that from other pieces of information: given a
prior probability distribution for the source redshift and galaxy
type, BPZ calculates the probability of the source being at redshift
$z_{s2}$ given its colors ${\bf C}$ and magnitude $m$, $P(z_{s2}|{\bf C},m)$. The stellar templates
used for the SED fitting are described by \citet{Coe}.  The F814W
magnitude is taken from \citet{Gavazzi}, where the brightness
distribution of the source was reconstructed after a lens
modeling. The value adopted is therefore $m_{\rm{F814W}} =
27.01\pm0.19$

\begin{figure}
\includegraphics[width = \columnwidth]{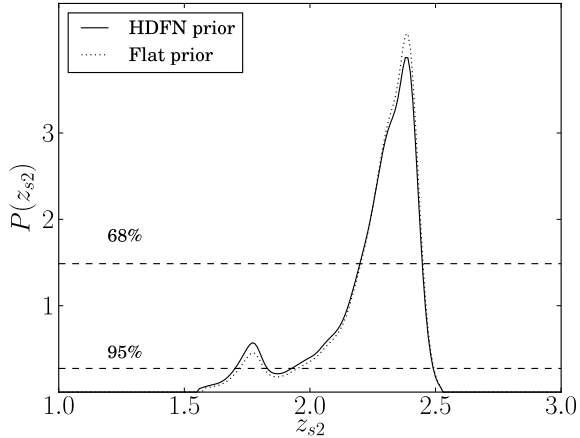}
\caption{{\em Solid line:} posterior probability distribution function of the source redshift, as calculated with BPZ, assuming a prior on $z_{s2}$ from Hubble Deep Field North number counts.
Overplotted are the levels corresponding to 68\% and 95\% enclosed probability.
{\em Dotted line:} posterior PDF assuming a flat prior on $z_{s2}$.
}
\label{PhotozPDF}
\end{figure}

For the redshift distribution we use a prior $P(z|m_{\rm{F814W}})$ suggested by \citet{Benitez} and based on number counts from the Hubble Deep Field North (HDFN).
Figure \ref{PhotozPDF} shows the redshift posterior probability distribution function $P(z_{s2}|{\bf C},m_{\rm{F814W}})$.
The most likely redshift with 68\% confidence interval is $z_{s2} = 2.41_{-0.21}^{+0.04}$.
As will be shown later, this information is sufficient to put interesting constraints on the model of the lens system.
We also calculated the photo-z assuming a flat prior on $z_{s2}$, and found a nearly identical result.
Colors calculated with a different lens light subtraction, the double-S\'{e}rsic model, yield the same photo-z well within the quoted uncertainties.

\section{Keck Spectroscopy}

The data were collected during the nights of 2006 December 23 and 24
with the LRIS instrument at the Keck Telescope I.  The original goals
of the observations were to measure a velocity dispersion profile of
the foreground deflector and to measure the redshift of the outer
ring.  The first goal was succesfully achieved, while we were not able
to detect any spectroscopic signature from the farthest source.

\begin{figure*}[!]
  \begin{center}
    \includegraphics[width=\textwidth]{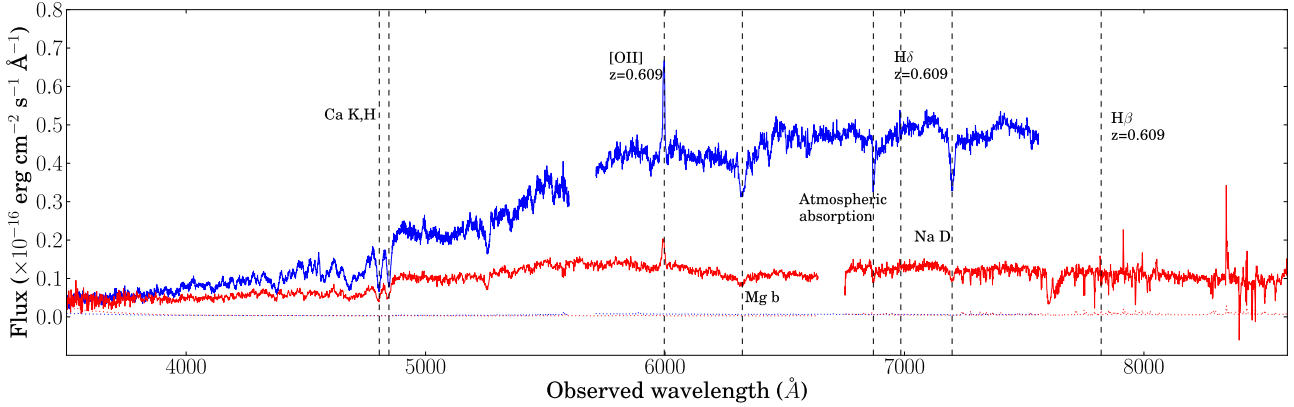}\\
  \end{center}
  \caption{LRIS spectra of the Jackpot. {\em Blue:} data from the first night. {\em Red:} data from the second night. The two spectra are extracted from rectangular apertures $1''\times3.36''$.
  Dotted line: noise level.}
  \label{Spectrum}
\end{figure*}

Because of the dual scope of our study, two different instrumental setups were used.
The first setup, used during the first night, was optimized for a better measurement of the velocity dispersion of the deflector.
The wavelength range in the red detector, the one used for the measurement of $\sigma$, was $\sim5700-7600\AA$, bracketing important absorption features in the rest frame of the lens at $z=0.222$.
During the second night we centered the slit on the longest arc of the outer ring, and used a setup with a broader wavelength range, up to $\sim8600\AA$.
A summary of the observations, with specifications on the setups used, is provided in Table \ref{Setups}.
\begin{deluxetable*}{ccccccccc}
\tablecaption{Spectroscopic observations: summary}
\tablehead{
\colhead{Date} & Exp. time & \colhead{Slit width} & \colhead{Dichroic} & \colhead{Blue grism} & \colhead{Red grating} & \colhead{Red $\lambda_c$} & \colhead{Weather} & \colhead{Seeing}}
\startdata
12/23/2006 & 16200 & 1.0'' & 560 & 600/4000 & 831/8200 & $6819 \AA$ & Good & $0.8''$\\
12/24/2006 & 12600 & 1.0'' & 680 & 300/5000 & 831/8200 & $7886 \AA$ & Good & $0.8''$ \\
\enddata
\label{Setups}
\end{deluxetable*}
The spectrum of the system is shown in Fig. \ref{Spectrum}.
There is no evidence for the presence of emission lines from objects other than the foreground lens and the inner ring.
Given our measurement of the photo-z of the outer ring, we would expect Ly-$\alpha$ emission to fall around $\sim4150\AA$, but it cannot be identified in our spectrum.
We can put an upper limit of $\sim5\times10^{-18}\mbox{ erg cm}^{-2}\mbox{ s}^{-1}$ to the flux in Ly-$\alpha$ from the source.

\subsection{Velocity dispersion}
The velocity dispersion of the main lens is measured by fitting stellar templates convolved with a Gaussian velocity distribution to the observed spectrum.
This operation is carried out with a Monte Carlo Markov Chain approach, using a code developed by M. W. Auger, and described by \citet{Suyu2010}.
The rest frame wavelength range used for the fit is $5100-5850\AA$.
For the stellar templates we used linear combinations of nine spectra from the INDO-US library, corresponding to K,G,F and A stars.
The most prominent absorption feature in the wavelength range considered is Mgb ($5175\AA$).
However, we experienced difficulty in finding a good fit to both Mgb and the rest of the spectrum.
It is known that some galaxies have enhanced magnesium features in the spectrum that are not well reproduced in standard stellar templates \citep{Barth}.
For this reason we decided to mask the Mgb absorption line out of the fitted spectrum.

\begin{figure}
\includegraphics[width = \columnwidth]{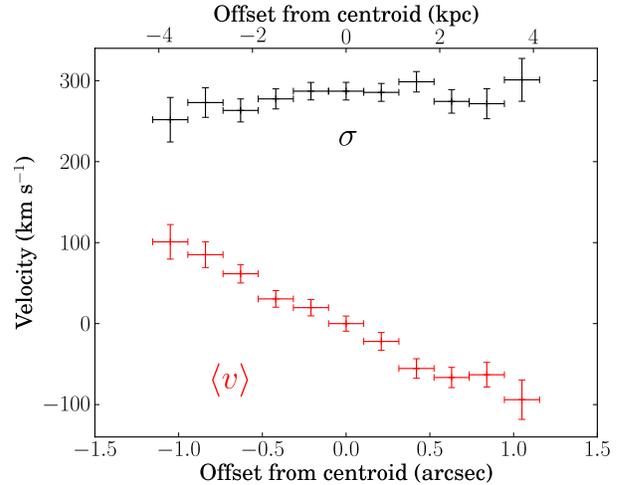}
\caption{Mean velocity and velocity dispersion profiles of the main lens within $1.15''$ from the centroid.}
\label{MeasProfile}
\end{figure}

\begin{figure}
\includegraphics[width = \columnwidth]{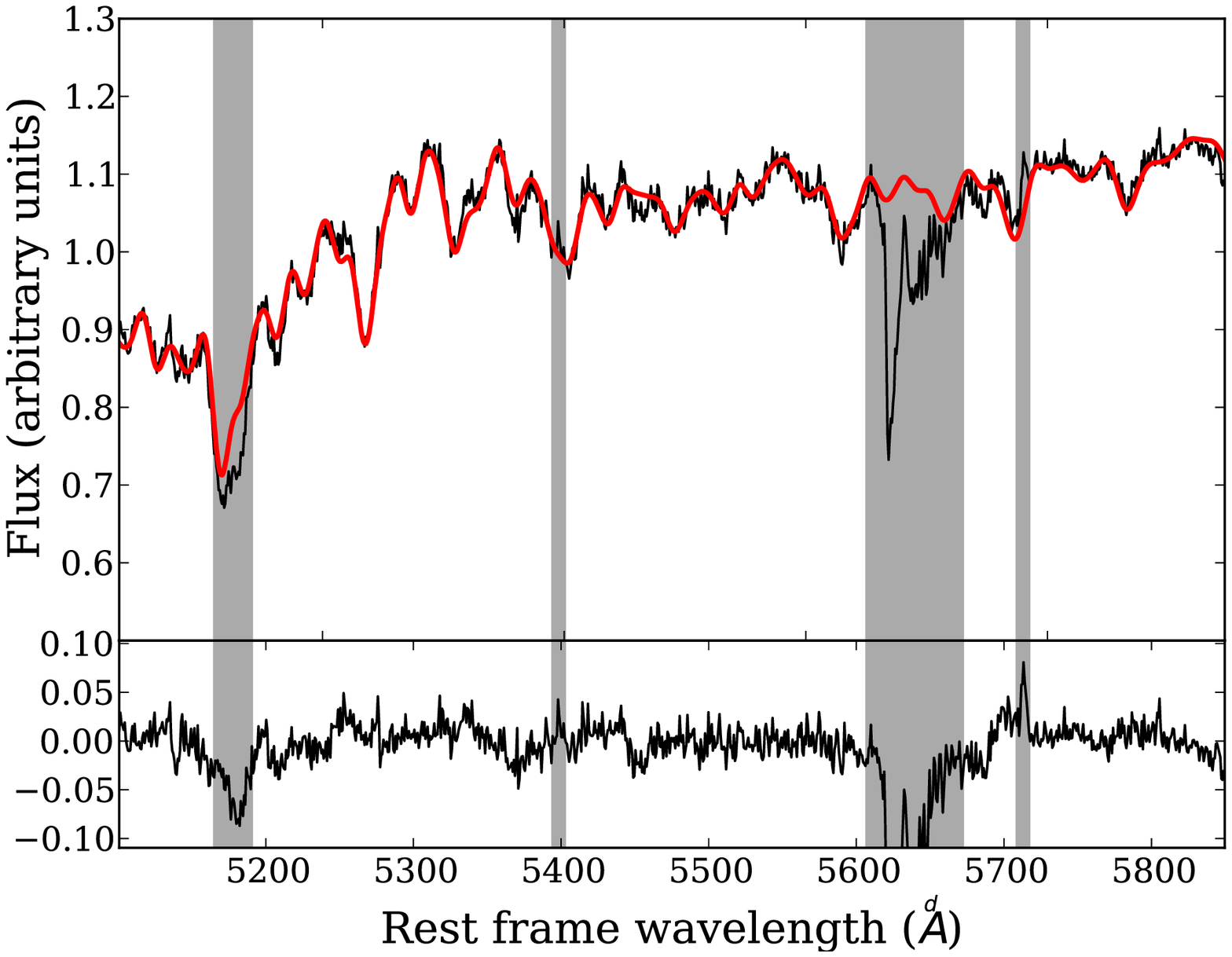}
\caption{Fit of the velocity dispersion of the lens. {\em Top:} The red curve is the best fit synthetic spectrum.
Shaded regions are masked and not used for the fit. {\em Bottom:} Residuals of the fit in fractions of the total flux.}
\label{FitSpectrum}
\end{figure}

With the aim of obtaining a velocity dispersion profile, we measured
$\sigma$ in a set of apertures.  The spatial position of the apertures
was determined by fitting the centroid of the trace of the lens in the
twodimensional spectra and it is accurate to $\sim0.02''$.  In Table \ref{vprofile-data} and Fig. \ref{MeasProfile} we report the measured values of $\sigma$ and
of the mean velocity in each aperture, while in Fig. \ref{FitSpectrum}
we show the fit in the central $0.42''$ as an example.  There is
evidence for some rotation, with $v_{\rm{rot}}^2 \ll \sigma^2$.

\begin{deluxetable}{ccc}
\tablecaption{Velocity profile measurements}
\tablehead{
\colhead{Slit offset} & \colhead{$\left<v\right>$} & \colhead{$\sigma$} \\
\colhead{(arcsec)} & \colhead{(km s$^{-1}$)} & \colhead{(km s$^{-1}$)} 
}
\startdata
-1.05 & $101\pm21$ & $252\pm25$ \\
-0.84 & $85\pm16$ & $273\pm18$ \\
-0.63 & $62\pm11$ & $263\pm14$ \\
-0.42 & $30\pm10$ & $278\pm12$ \\
-0.21 & $20\pm10$ & $287\pm11$ \\
0.00 & $0\pm9$ & $287\pm11$ \\
0.21 & $-22\pm11$ & $286\pm11$ \\
0.42 & $-55\pm12$ & $299\pm13$ \\
0.63 & $-67\pm13$ & $274\pm15$ \\
0.84 & $-63\pm15$ & $272\pm19$ \\
1.05 & $-94\pm24$ & $301\pm25$
\enddata
\tablenotetext{}{Mean velocity and velocity dispersion profile. Apertures are $1.00\times0.21''$ rectangles.}
\label{vprofile-data}
\end{deluxetable}

\section{Stellar masses}\label{mstarsect}

Here we present a measurement of the stellar mass of the foreground
lens galaxy. The procedure adopted is the following: we fit stellar
population synthesis models to the observed spectral energy
distribution (SED) of the galaxy. A measurement of this kind was already
performed by \citet{Grillo2009} and \citet[Paper IX]{PaperIX} for the
same object. Their results agree within the errors. \citet{Grillo2009}
used SDSS multiband photometry ($u$, $g$, $r$, $i$, $z$ bands) as
their observed SED. In Paper IX, high resolution HST data was used,
but only in two bands (F814W and F606W). \citet{PaperIX} also
introduced a powerful statistical analysis method, based on Bayesian
statistics that allows for physically meaningful priors on the model
parameters as well as a full exploration of uncertainties and
correlation between the inferred parameters.  With five band HST
photometry we can now extend the analysis of Paper IX, to obtain a
more robust estimate of the stellar mass.

The fitting method is the same as that developed by \citet{PaperIX}, and can be summarized as follows. Composite stellar population models are created from \cite{BC03} stellar templates. The star formation history is modeled with a single exponentially decaying burst.
The parameters of the model are age, metallicity, exponential burst timescale, dust reddening and stellar mass.
The parameter space is explored using a Monte Carlo Markov Chain (MCMC) routine, through which the posterior PDF is characterized.
The stellar templates used are based on either a Salpeter or a Chabrier IMF.
For the description of the photometry of the lens we use the double tPIEMD model described in Sect. \ref{LensPhoto}, that is consistent with the analyses presented in the following Sections. The stellar masses of the two components are fitted independently.
Results are listed in Table \ref{mstartable}, together with the values previously found by \citet{Grillo2009} and \citet{PaperIX}.

The analysis reveals the presence of dust for component 1, coherently with our previous findings.
Repeating the fit with the dust-corrected magnitudes yields indistinguishable stellar masses.
The logarithm of the stellar masses changes by 0.06 if we use the description of the light profile with S\'{e}rsic components instead of tPIEMDs.
This is due to the different behavior at large radii of the two profiles.
Differences in the mass within the outer Einstein radius for the two models are instead well within the measurement errors.

\begin{deluxetable}{cccc}
\tablecaption{Stellar mass of the foreground lens, from SPS models}
\tablehead{
\colhead{IMF} & \colhead{Chabrier} & \colhead{Salpeter} & \\
 & $\log(M_*/M_\Sun)$ & $\log(M_*/M_\Sun)$ & }
\startdata
Comp. 1 & $10.85_{-0.06}^{+0.09}$ & $11.13_{-0.11}^{+0.05}$ & This work\\
Comp. 2 & $11.27_{-0.08}^{+0.05}$ & $11.52_{-0.08}^{+0.06}$ & This work\\
Total & $11.40 \pm 0.06$ & $11.66 \pm 0.06$ & This work\\
 & $11.38_{-0.12}^{+0.04}$ & $11.61_{-0.08}^{+0.02}$ & \citet{Grillo2009} \\
 & $11.34 \pm 0.12$ & $11.59 \pm 0.12$ & \citet{PaperIX} \\
\enddata
\label{mstartable}
\end{deluxetable}

\section{A single component model: measuring the average slope}\label{LensingSect}

In this Section we present a single-component lensing and dynamics study of the foreground galaxy, where the total density distribution of the lens is described with a power-law.
The goal is to obtain a measurement of the slope of the total mass profile and also to test the accuracy allowed by our data in constraining mass models.
The system, with its two Einstein rings, offers more constraints than typical single-source lenses.
However, the analysis is complicated by the presence of two different lenses along the line of sight.
Light rays from the second source are first deflected by the object corresponding to the inner ring and then by the foreground lens, with the result that, unlike the single lens case, the relation between the size of the outer Einstein ring and the enclosed projected mass of the lens is nontrivial.
Nevertheless, this can be properly accounted for as described below.

A first lens modeling of the system was carried out in Paper VI. 
The procedure adopted there was a conjugate points method: multiply imaged spots in the lensed features are identified, and the lens model is determined by minimizing the distance between the corresponding points in the source plane. This is a conservative approach, since it does not make use of all of the information from the surface brightness of the rings.
The main lens was modeled as a power law ellipsoid, with dimensionless surface mass density $\kappa \equiv \Sigma/\Sigma_{cr}$ given by:
\begin{equation}
\kappa(\vec{r},z_s) = \frac{b_\infty^{\gamma'-1}}{2}(x^2 + y^2/q^2)^{(1-\gamma')/2}\frac{D_{\rm{ls}}}{D_{\rm{os}}},
\end{equation}
where $b_\infty = 4\pi(\sigma_{\rm{SIE}}/c)^2$ and $D_{\rm{ls}}$ ($D_{\rm{os}}$) is the angular diameter distance of the source relative to the lens (observer).
The second lens
(first source corresponding to the brighter arc) was modeled as a
singular isothermal sphere (SIS).  The model parameter space was
explored via a MCMC.  The results showed
that two types of solution are possible: a model with larger $\sigma_{\rm{SIE}}$, shallower slope $\gamma'$ and less massive second lens, or a model with a more massive second lens and steeper main lens slope
(see Figure 9 of Paper VI, or black contours of
Figure \ref{contourplots}). Part of this degeneracy was due to our
ignorance of the redshift of the outer ring.

In this Paper we use the lens model of Paper VI described above and improve it by incorporating 1) our measurement of photo-z of the outer ring and 2) a stellar dynamics analysis.

\subsection{Stellar dynamics modeling}\label{Jeanssect}

We wish to use our measurements of the velocity dispersion profile of the lens to constrain our lens models.
This is done with a procedure similar to that adopted by \citet{Suyu2010}, which can be described as follows.
For a given model provided by the lensing analysis, we compute a model velocity dispersion profile and compare it to the observed one.
The model velocity dispersion is obtained by solving the spherical Jeans equation
\begin{equation}\label{Jeans}
\frac{1}{\rho_*}\frac{d\rho_*\sigma_r^2}{dr} + 2\frac{\sigma_\theta^2}{r} = -\frac{GM(r)}{r^2},
\end{equation}
where $\rho_*(r)$ is the density distribution of the light, $\sigma_r$ and $\sigma_\theta$ are the radial and tangential components of the velocity dispersion tensor, $M(r)$ is the total mass enclosed within the spherical shell of radius $r$.
We impose spherical symmetry in the mass model by adopting a spheroidal mass distribution
\begin{equation}
\rho(r) \propto r^{-\gamma'}
\end{equation}
with normalization chosen such that the total projected mass enclosed within the Einstein radius equals that of the corresponding circularized lens model.
The light distribution is described as the sum of two tPIEMD profiles, with the same parametrization described in Section \ref{LensPhoto} (best-fit parameters are in table \ref{twotNIEfit}).
The 3d stellar distribution corresponding to the surface brightness profile (\ref{tNIEsb}) used to fit the photometry is
\begin{equation}\label{rhotNIE}
\rho(r) = \rho_cr_c^2\left[\frac{1}{r_c^2+r^2} - \frac{1}{r_t^2+r^2}\right],
\end{equation}
with $r \equiv x^2/q_* + q_*y^2 + z^2$.
Here we set the axes ratios $q_*$ to one, as we are assuming spherical symmetry.

We then assume a Osipkov-Merritt model for the velocity dispersion tensor \citep{Osipkov,Merritt}:
\begin{equation}
\frac{\sigma_{\theta}^2}{\sigma_r^2} = 1 - \frac{r^2}{r_a^2+r^2},
\end{equation}
where $r_a$ is the anisotropy radius (orbits are radially anisotropic beyond $r_a$).
Finally, we simulate the line-of-sight velocity dispersion measured in our apertures.
Rotation is neglected.
Although the lens is seen to be rotating, its mean velocity is small compared to the velocity dispersion and should not contribute much to the dynamics of the object.
The effect of this approximation will be discussed further below.

\subsection{Combining the constraints}\label{importancesect}

The models of the lens are defined by the set of parameters $\boldsymbol{\eta} \equiv\{\sigma_{\rm{SIE,lens}}, \gamma', \sigma_{\rm{SIS,s1}}, z_{s2}\}$: the strength and power-law index of the foreground lens, the strength of the background lens and the redshift of the outer ring, respectively.
Each model gives a prediction of the velocity dispersion in each aperture, $\sigma_{\rm{ap},\mathit{i}}^{\rm{(mod)}}$.
The new posterior probability distribution for the model is obtained via importance sampling: the MCMC sample corresponding to the lens modeling of Paper VI is weighted by the likelihood of the measurements ${\bf d} \equiv \{z_{s2},\sigma_{\rm{ap},\mathit{i}}^{\rm{(meas)}}\}$ given the model parameters $\boldsymbol{\eta}$.
The following likelihood function is used:
\begin{equation}
L({\bf d}|{\boldsymbol{\eta}}) = P_z(z_{s2})\prod_iG_i(\sigma_{\rm{ap},\mathit{i}}^{\rm{(meas)}}|{\boldsymbol{\eta}})
\end{equation}
where $P_z(z_{s2})$ is the PDF in Figure \ref{PhotozPDF} and
\begin{equation}
G_i(\sigma_{\rm{ap},\mathit{i}}^{\rm{(meas)}}|{\boldsymbol{\eta}}) = \frac{1}{\sqrt{2\pi\Delta_{\sigma,\mathit{i}}^2}}\exp{-\frac{(\sigma_{\rm{ap},\mathit{i}}^{\rm{(meas)}}-\sigma_{\rm{ap},\mathit{i}}^{\rm{(mod)}})}{2\Delta_{\sigma,\mathit{i}}^2}},
\end{equation}
and $\sigma_{\rm{ap},\mathit{i}}$ and $\Delta_{\sigma,\mathit{i}}$ are the zeroth and second moment of the posterior PDF of the measured velocity dispersion in aperture $i$, respectively.

\begin{figure}
\begin{tabular}{c}
\includegraphics[width = \columnwidth]{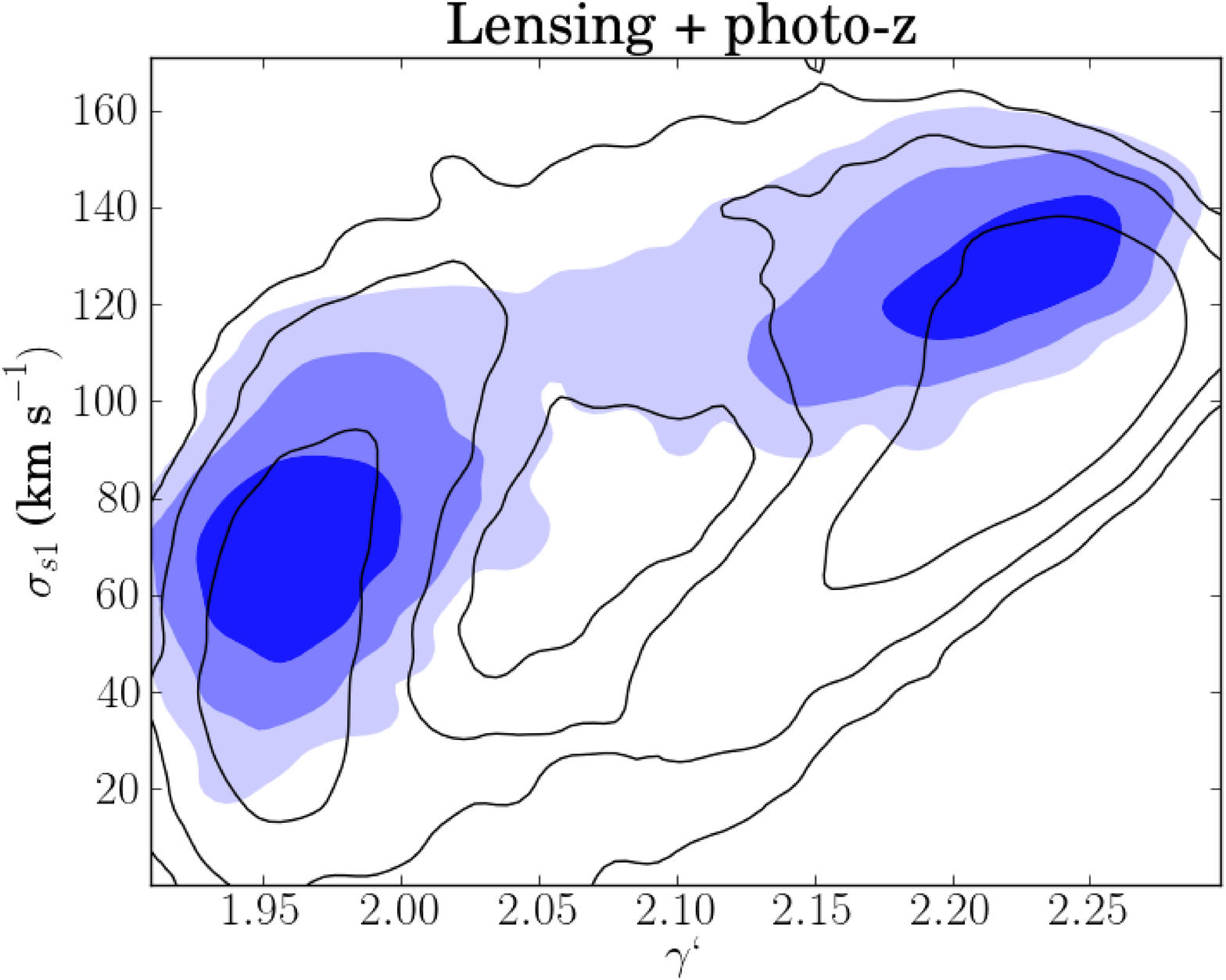} \\
\includegraphics[width = \columnwidth]{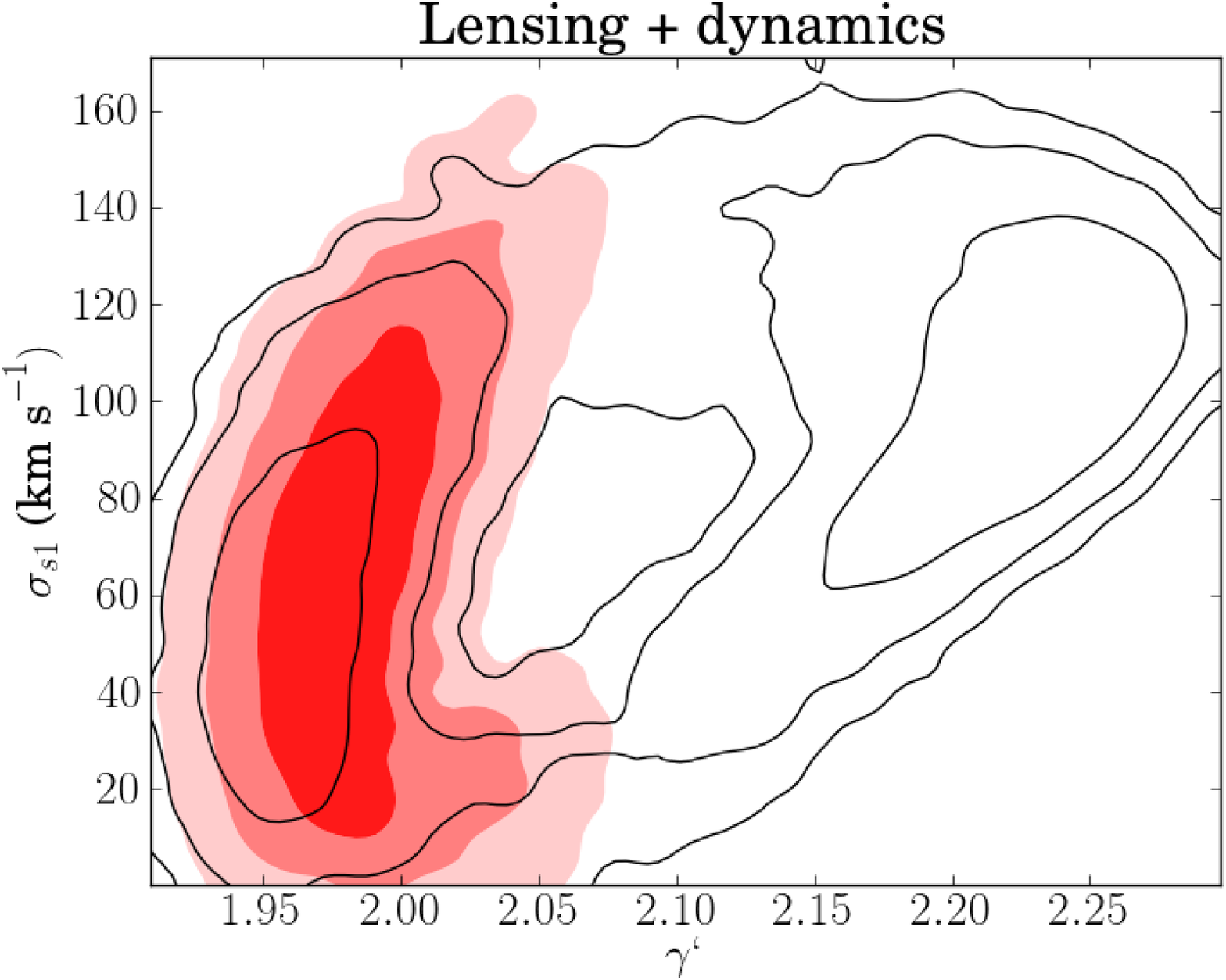} \\
\includegraphics[width = \columnwidth]{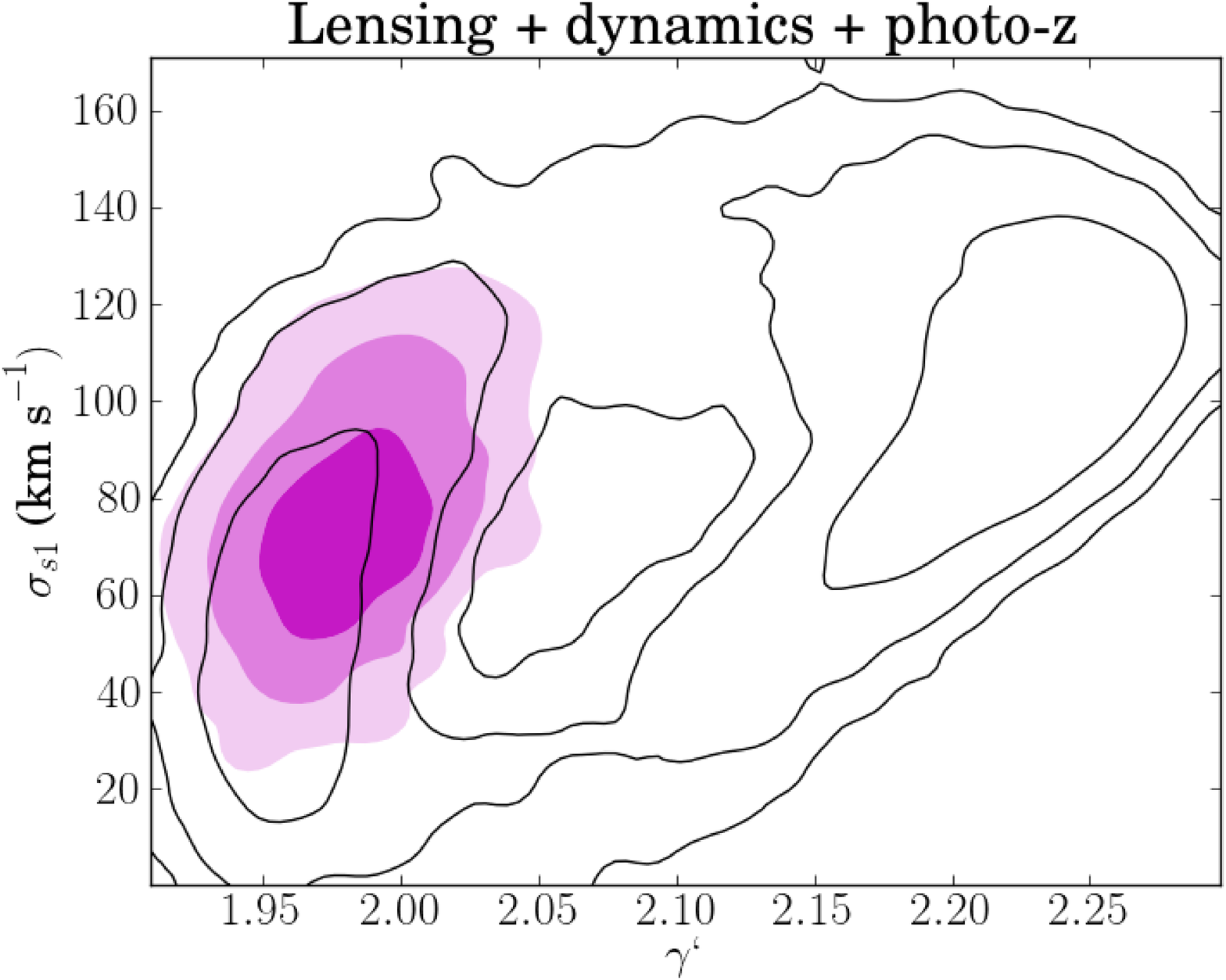}
\end{tabular}
\caption{Posterior PDF of $\gamma'$ and $\sigma_{s1}$ of the updated (filled contours) lens model, together with the old model of Paper VI (empty contours).
The updated model includes only the photo-z measurement of the outer ring in the top panel, only the velocity dispersion profile of the lens in the middle panel, and both the photo-z and velocity dispersion profile in the bottom panel.
The levels correspond to 68\%, 95\% and 99.7\% enclosed probability.}

\label{contourplots}
\end{figure}

In Fig. \ref{contourplots} we show the updated Posterior PDF obtained
by importance sampling with the photo-z and dynamics measurements,
both separately and jointly.  It is clear that although photo-z and
stellar kinematics alone leave some degeneracies, the posterior pdfs
are almost perpendicular in this space, and therefore the combination
of the two is particularly effective. 
The estimate of the slope obtained by marginalizing over the other parameters is
\begin{equation}\label{bestgamma}
\gamma' = 1.98\pm0.02.
\end{equation}
We stress that our uncertainty on this parameter is a factor of four smaller than the typical error on $\gamma'$ from studies of single-source gravitational lenses with SDSS spectroscopy \citep[see Figure \ref{comparison}]{PaperX}. Comparable precision was reached by \citet{Bar++11} for a sample of lens systems with two dimensional kinematics constraints from integral field spectroscopy.

In order to better understand the significance of these results, we try to quantify the error introduced by our simplified model for the stellar dynamics.
Two of our assumptions are potential sources of bias: spherical symmetry and the non-rotating approximation. The uncertainty in the mass determination from kinematics data is of order $\delta\sigma^2/\sigma^2\sim10\%$. 
Biases on the order of this uncertainty or smaller are unlikely to bring significant changes to the results of our analysis.
By considering only the velocity dispersion and neglecting rotation, we underestimate the mass of the galaxy by a factor $\sim (v_{\rm{rot}}/\sigma)^2$, which is within $10\%$ in all apertures but one.
To gauge the importance of this effect we perform the following test.
We fit the model velocity dispersion profiles to the following ``effective velocity dispersion'': $\sigma_{\rm{eff}} \equiv \sqrt{\sigma^2 + v_{\rm{rot}}^2})$.
We then apply the same importance sampling procedure described above to get a new constraint on the density slope $\gamma'$. The new estimate with $1\sigma$ uncertainty is:
\begin{equation}
\gamma' = 1.97\pm0.02,
\end{equation}
which is consistent with the original estimate given by (\ref{bestgamma}). On the basis of this result, we can conclude that our approximation of non-rotating halo introduces a systematic error of order 0.01 on the inferred value of the slope $\gamma'$.

Quantifying the systematics introduced by the spherical symmetry assumption is more complicated.
In a previous work, \citet{Bar++11} performed a robust dynamical modelling of 12 SLACS lenses previously analysed with a spherical Jeans equation approach by \citet{PaperX}. The slopes $\gamma'$ inferred by \citet{PaperX} are consistent with the more accurate measurements of \citet{Bar++11}, with a bias on $\gamma'$ of $0.05\pm0.04$.
However, the uncertainty on $\gamma'$ that we achieve in our work is smaller than that and an estimate of the bias requires additional work.
Two distinct effects come into play.
First, the lens has a non-circular projected shape in both its mass and light distribution.
Second, the galaxy may even have asymmetries along the line of sight.
The importance of these effects on our analysis is quantified in Appendix B.
By relaxing the assumption of spherical symmetry the additional uncertainty on the velocity dispersion is about $\delta\sigma^2/\sigma^2\sim10\%$.
It follows that none of our results change appreciably.

An independent analysis of the system was carried out by \citet{Vegetti}. The method adopted by them is more complex than the one used in Paper VI: they made use of information from all the pixels of the lensed features to reconstruct the source surface brightness as a whole.
Using data from the inner ring only, they obtained the following estimate for the density slope:
\begin{equation}\label{gammaSimona}
\gamma' = 2.20\pm0.03^{\rm{(stat)}}.
\end{equation}
This is a local estimate of the slope $\gamma'$, obtained by measuring the magnification of the arc in the radial direction.
Our measurement is instead an average slope, obtained by fitting a single power-law halo to data spanning the lens from the center (dynamics) to the outer lensed ring.
This difference may suggest that the actual mass distribution of the lens is different from a simple power-law halo.
It is also for this reason that we proceeded to model the system with a more complex model.

\section{A two-component analysis: dissecting luminous and dark matter}\label{twocompsect}

We perform a two-component lensing and dynamics study where the mass distribution is composed of a dark matter halo and a bulge of stars.

\subsection{Lensing and dynamics modeling}
We use a power-law ellipsoid for the dark matter, while the stars are described with the double tPIEMD model found from the photometry analysis. 
The second lens is again modeled as a SIS.
The parameters of the stellar distribution are fixed to the best-fit values reported in Table \ref{twotNIEfit}.
The global mass-to-light ratio is left as a free parameter, but the relative contribution of the two components is fixed according to the results of the stellar population synthesis analysis presented in Section \ref{mstarsect}.
For a unit F814W-band magnitude, component 1 is measured to be a factor of 1.73 (1.77) heavier than component 2 assuming a Salpeter (Chabrier) IMF.
In our lensing model, the mass-to-light ratio of component 1 is set to be 1.75 times larger than for component 2.

We also allow for constant external shear $\gamma_{\rm{ext}}$ with position angle $\rm{PA}_{\rm{ext}}$ and constant external convergence $\kappa_{\rm{ext}}$ in the lens plane.
Issues related to the external convergence are discussed below in a dedicated subsection.
Compared to the lensing study presented in the previous section, this model has two additional free parameters: the stellar mass $M_*^{\rm{LD}}$ and the external convergence $\kappa_{\rm{ext}}$.
Given the very tight constraint on the average slope $\gamma'$ from the single component analysis, we expect to be able to determine both the slope of the dark matter halo $\gamma_{\rm{DM}}$ and the stellar mass $M_*^{\rm{LD}}$ with sufficient accuracy.
The range of values of the slope of the dark matter halo explored in this analysis is $1.0 < \gamma_{\rm{DM}} < 3.0$. 

The technique adopted to fit the model to the lensing data is the same
used for Paper VI: a conjugate points method implemented with a
MCMC.  The dynamics analysis is carried out with a
procedure very similar to the one described in \S~\ref{Jeanssect}: we
solve the spherical Jeans equation for our model and obtain a
synthetic velocity dispersion profile to be compared to the measured
one.  The (spherically symmetric) model mass distribution is obtained
by circularizing the projected mass distribution of the lens model,
setting $q_{\rm{DM}}$ and $q_*$ to one, and by taking the
corresponding spherical deprojections.  The light distribution is set
by circularizing the double tPIEMD profile specified in Table
\ref{twotNIEfit}.

We then proceed to incorporate information on stellar dynamics and on the redshift of the background source.
This is done by importance sampling, with the same method described in \S~\ref{importancesect}.

\subsection{External convergence}

Objects other than the main lens can contribute to the surface mass
density $\kappa$.  This external convergence is hard to detect and is
degenerate with the total mass of the lens galaxy. Ignoring the
contribution to $\kappa$ from perturbers can lead in principle to a
bias in the measurement of the key parameters of the lens.  In order
to take into account the effect of external convergence on our error
budget, we include it in our model by generating random values of
$\kappa_{\rm{ext}}$ drawn from a plausible distribution.
This procedure allows us to propagate correctly this uncertainty to the other model parameters.
Kinematics information can also help to constrain $\kappa_{\rm{ext}}$ to some extent, as it is only sensitive to the mass dynamically associated with the galaxy, in contrast to lensing that is sensitive to all mass structures along the line of sight to the source.

Insight on the actual value of $\kappa_{\rm{ext}}$ can be gained by studying the lens environment.
According to \citet{PaperVIII}, this is found to be marginally underdense with respect to average lines of sight, therefore there is no evidence for the presence of a group in the lens neighborhood.
The closest cluster known to the NASA/IPAC Extragalactic Database (NED) is MaxBCGJ146.87912+10.07800, at redshift $z=0.151$ and projected distance 8.70 arcmin from our lens \citep{PaperVIII}.
If we assume a SIS profile for the cluster with a typical value for its velocity dispersion $\sigma = 1000\mbox{ km s}^{-1}$ we obtain a contribution to the convergence $\kappa_{\rm{cl}} < 0.01$.
We also scanned the Sloan Digital Sky Survey archive looking for massive red galaxies within 5' of the lens.
Only one early-type galaxy was found, at a redshift $z=0.218$ and angular distance 2.6'.
If we assume that this object is the brightest galaxy of a group and associate it with a SIS halo of $\sigma = 500\mbox{ km s}^{-1}$ the corresponding convergence at the location of the lens is $\kappa=0.02$.
Finally, the lensing analysis of \citet{Gavazzi} quantified the external shear as $\gamma_{\rm{ext}} = 0.07$ directed $-31$ degrees East of North.
The HST images show two objects with the same alignment relative to the lens (see Fig. \ref{IRtidal}).
If we make the assumption that those objects are responsible for the shear and assume again a SIS profile we obtain $\kappa_{\rm{ext}} = |\gamma_{\rm{ext}}| = 0.07$.

\citet{Hilbert} studied the external convergence associated with strong lensing systems in cosmological simulations.
They found that for a source at redshift $z_s = 5.7$ the distribution of $\kappa_{\rm{ext}}$ is skewed with a peak at $-0.04$, has zero mean and a scatter of $0.05$.
A slightly smaller scatter and a peak at $-0.02$ is found by \citet{Suyu2010} for sources at $z_s = 1.39$.

Taking all these aspects into account, we adopt as prior for $\kappa_{\rm{ext}}$ in our analysis a Gaussian distribution peaked at $0.05$, with dispersion $\sigma_\kappa = 0.05$ and truncated to values in the interval $-0.05 < \kappa_{\rm{ext}} < 0.15$.
This range should capture the indication of a positive contribution from the object responsible for the shear and take into account the effect of random mass clumps along the line of sight.
Priors with a broader range of allowed values of $\kappa_{\rm{ext}}$ lead to larger uncertainties on the other model parameters, but none of the conclusions of our study is altered.

\subsection{Results}

\begin{figure*}
\includegraphics[width = \textwidth]{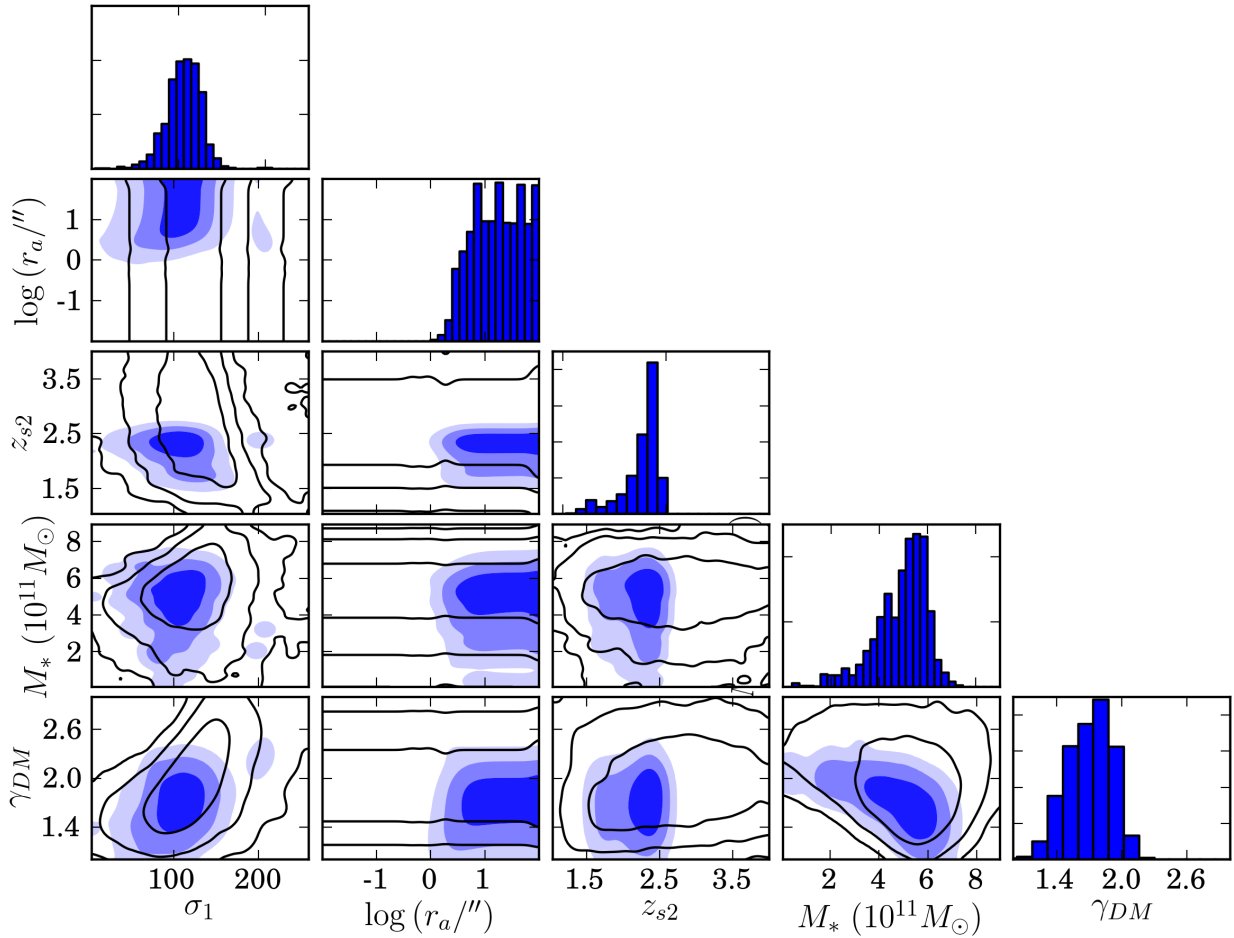}
\caption{Posterior PDF in the multidimensional space spanned by the stellar mass $M_*^{\rm{LD}}$, slope of the dark matter halo $\gamma_{\rm{DM}}$, radial anisotropy scale radius $r_a$, strength of the second lens $\sigma_{s1}$ and redshift of the second source $z_{s2}$.
The levels correspond to 68\%, 95\%, 99.7\% enclosed probability.
{\em Solid contours:} constraints from lensing only. {\em Shaded regions:} constraints from lensing, dynamics, and photo-z.}
\label{PDFtwocomp1}
\end{figure*}

\begin{figure*}
\includegraphics[width = \textwidth]{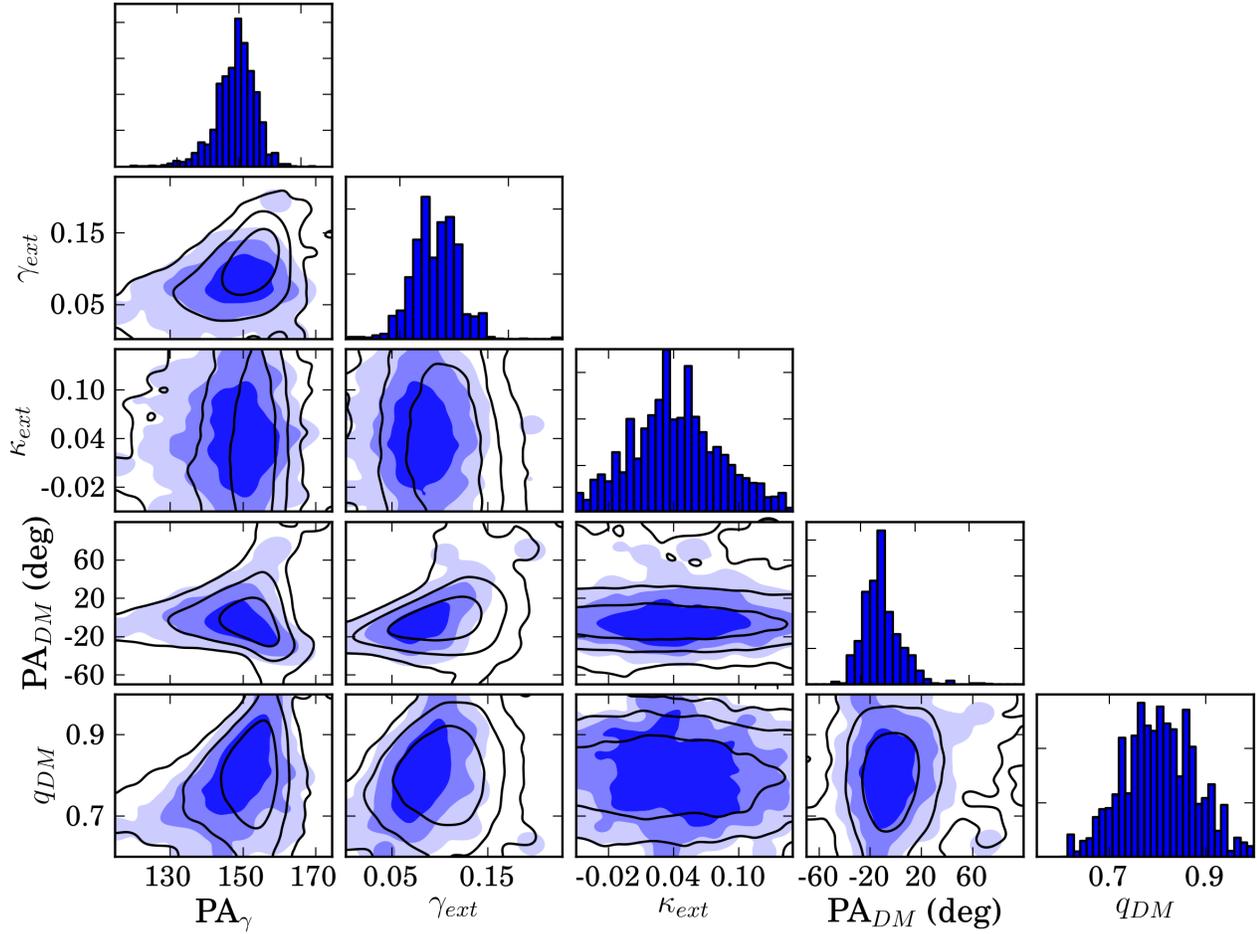}
\caption{Posterior PDF in the multidimensional space spanned by external convergence $\kappa_{\rm{ext}}$, strength and position angle of the external shear, $\gamma_{\rm{ext}}$, $\rm{PA}_{\rm{ext}}$, axis ratio of the dark matter halo $q_{\rm{DM}}$, position angle of the major axis of the dark matter halo, $\rm{PA}_{\rm{DM}}$.
The levels correspond to 68\%, 95\%, 99.7\% enclosed probability.
{\em Solid contours:} constraints from lensing only. {\em Shaded regions:} constraints from lensing, dynamics, and photo-z.}
\label{PDFtwocomp2}
\end{figure*}

\begin{figure}
\includegraphics[width = \columnwidth]{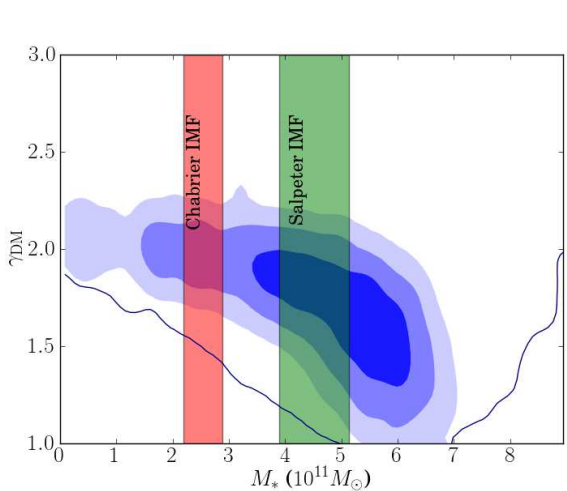}
\caption{
Posterior PDF projected in the space $M_*-\gamma_{\rm{DM}}$. The vertical shaded regions show independent measurements of the stellar mass from photometry, presented in Section 5.}
\label{Mstargamma}
\end{figure}

\begin{figure}
\includegraphics[width = \columnwidth]{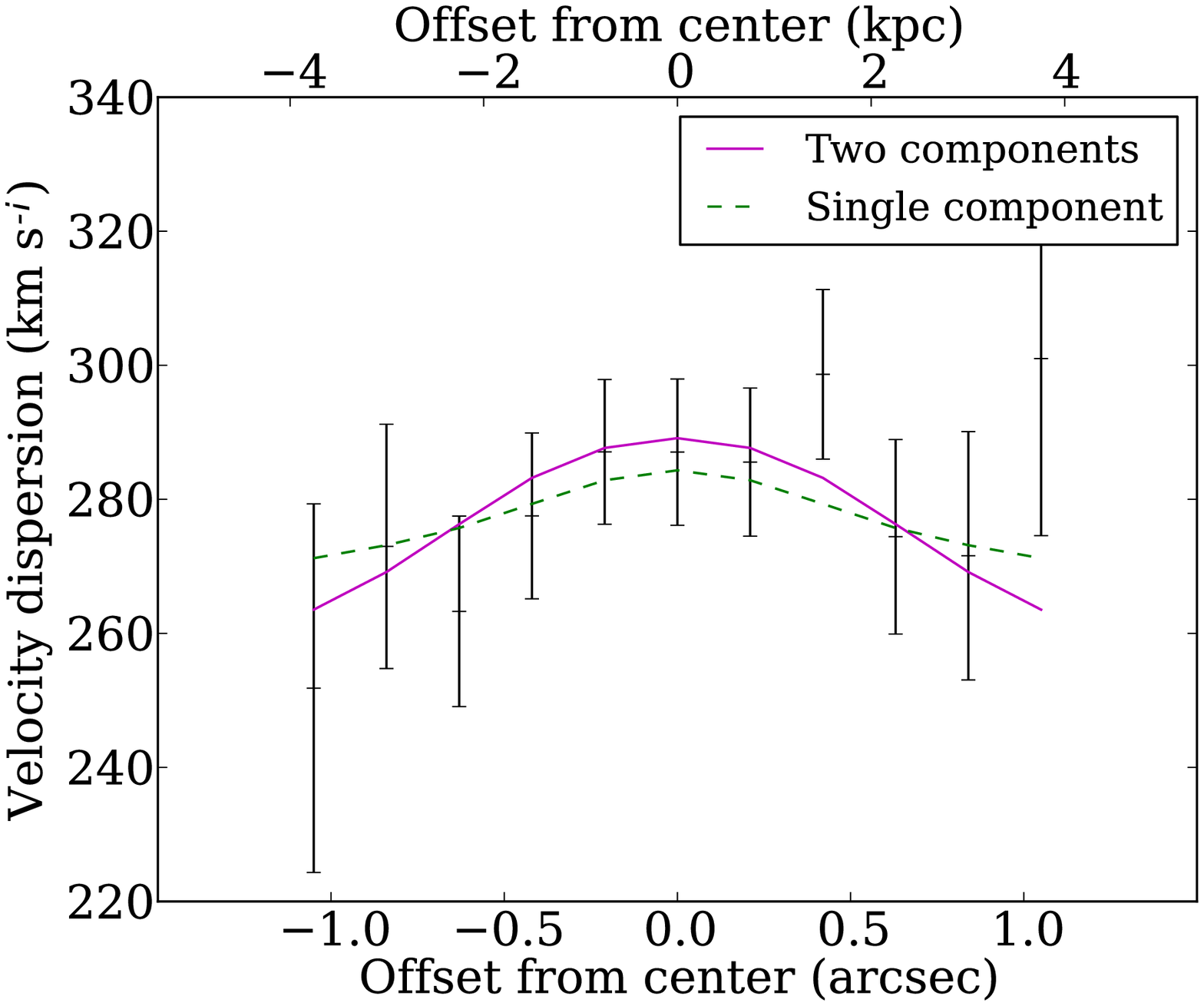}
\caption{Best-fit velocity dispersion profile of the lens.
{\em Solid line:} two components model. {\em Dashed line:} single power-law model.}
\label{sigmaprofiles}
\end{figure}

Contour plots of the posterior PDF for
the model parameters are shown in Figures \ref{PDFtwocomp1} and
\ref{PDFtwocomp2}.  The best-fit velocity dispersion profile is
plotted in Figure \ref{sigmaprofiles}.
The inference on the two key parameters $M_*$ and $\gamma_{DM}$ is shown in better detail in Figure \ref{Mstargamma}.
By marginalazing over the remaining parameters, our model constrains the stellar mass to
\begin{equation}
M_* = 5.5_{-1.3}^{+0.4}\times10^{11}M_\Sun.
\end{equation}
This estimate comes from lensing and dynamics data, and does not rely
on assumptions on the mass-to-light ratio of the stars.  This value
will be compared with the measurement of the stellar mass obtained
independently from photometry.

\begin{figure}
\begin{tabular}{c}
\includegraphics[width = \columnwidth]{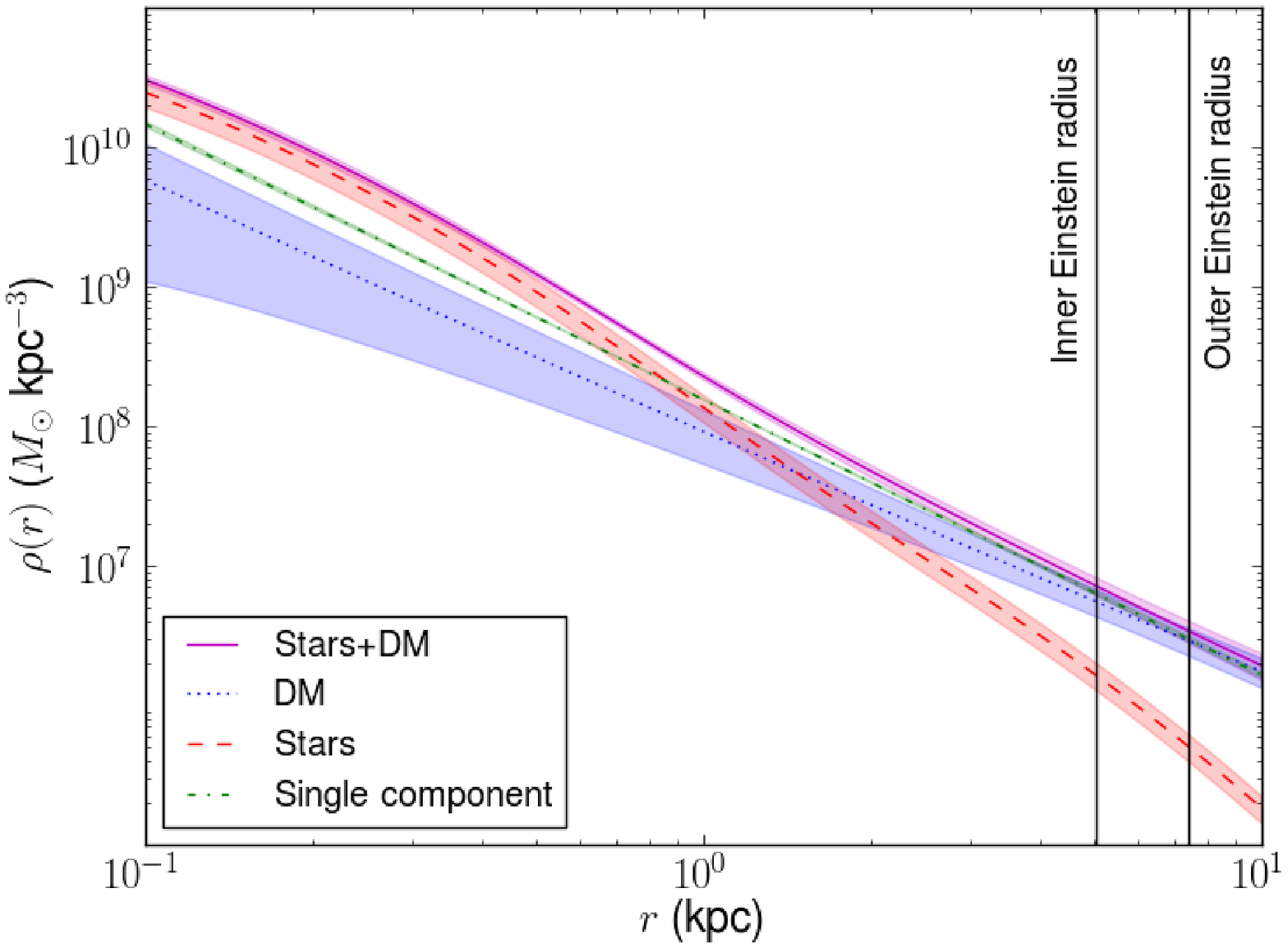} \\
\includegraphics[width = \columnwidth]{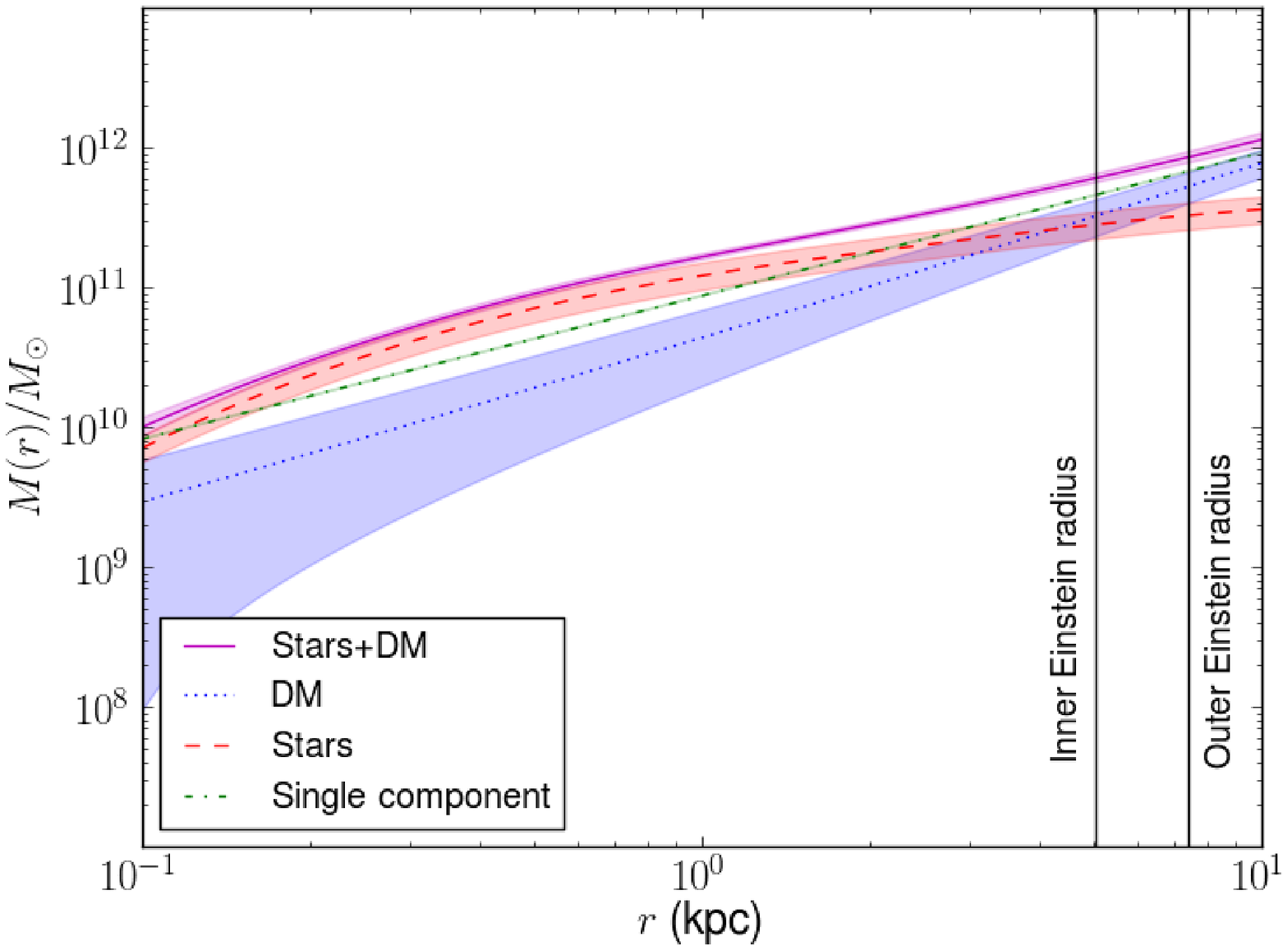}
\end{tabular}
\caption{Best-fit density (top) and mass (bottom) profiles. {\em Solid line:} total mass from bulge-halo decomposition. {\em Dashed line:} stellar mass. {\em Dotted line:} dark matter. {\em Dash-dotted line:} total mass from single component analysis. The shaded regions represent $1-\sigma$ uncertainties.}
\label{profiles}
\end{figure}

Another important result is the constraint that we obtain on the slope of the dark matter halo:
\begin{equation}
\gamma_{\rm{DM}} = 1.7\pm0.2
\end{equation}
This result shows strong evidence for a contraction of the dark matter distribution relative to the $r^{-1}$ inner slopes typical of dark matter only simulations \citep{NFW}.
Figure \ref{profiles} shows the mean density profile of each mass component compared to the mean single power-law fit from Sect. \ref{LensingSect}.

Our inference for the anisotropy radius constrains $r_a > 13$ kpc, meaning that radial anisotropy is ruled out in the region probed by our data.
This
is consistent with previous work \citep[e.g.][]{K+T03,TreuKoop} and
expected on theoretical grounds, because strong radial anistropy would
lead to instabilities.

In contrast to the power-law model considered in the previous section,
this mass model has a density distribution with a slope that changes
with radius.  It is interesting to compare the local value of the
slope at the location of the inner ring with the measurement of
\citet{Vegetti}. \citet{Vegetti} modeled the HST F814W image using only lensing information from the inner ring.  Lensing is
only sensitive to projected masses, therefore, in order for the
comparison to be meaningful, we have to consider the logarithmic slope
of the total projected mass distribution, evaluated at the inner
Einstein radius. We find
\begin{equation}
\frac{d\log{\kappa}}{d\log{r}} = -1.1\pm0.1.
\end{equation}
This value is consistent with the slope found by \citet{Vegetti}, which is given by $-(\gamma'-1) = -1.2$, where $\gamma'$ is the slope of the 3d mass distribution given in (\ref{gammaSimona}).

Finally, it is interesting to note how the inference on the stellar
mass is rather insensitive to the actual value of the redshift of the
second source, $z_{s2}$ (see Figure \ref{PDFtwocomp1}).  This means
that, with the current data quality, a spectroscopic measurement of
the redshift of the outer ring would not bring significantly more
information.

%


\section{Discussion}\label{IMFsect}

\subsection{Luminous and dark matter in the lens}

The data in our possession allowed us to study the lens galaxy of the system SDSSJ0946+1006 under multiple aspects.
Thanks to the high resolution photometry from HST we were able to note now the light distribution is well described with two components, while single component models yield poor fits.
These two components appear to be nearly perpendicular (in projection), have significantly different effective radii and surface brightnesses.
The colors of the more compact component (component 1 from now on) are also significantly redder (see Table \ref{colortable}), indicator of an older or more metal-rich stellar population.
As we will discuss below, these characteristics suggest a particular scenario for the past evolution of this object.

In Sections \ref{mstarsect} and \ref{twocompsect} we presented two independent measurements of the stellar mass of the foreground galaxy of the system SDSSJ0946+1006 derived with a lensing+dynamics analysis and with a stellar population synthesis study.
The measured values of $M_*$, obtained by marginalizing over the other model parameters, are reported in Table \ref{mstarmeas}.

\begin{deluxetable}{ccc}
\tablecaption{Stellar mass of the foreground galaxy}
\tablehead{
\colhead{Method} & \colhead{$M_*$ $(M_\Sun)$} & \colhead{$\alpha$\tablenotemark{a}}}
\startdata
Lensing+dynamics & $5.5_{-1.3}^{+0.4}\times10^{11}$ & \\
SPS, Chabrier IMF & $(2.5\pm0.3)\times10^{11}$ & $2.0\pm0.4$ \\
SPS, Salpeter IMF & $(4.5\pm0.6)\times10^{11}$ & $1.1\pm0.2$
\enddata
\tablenotetext{a}{$\alpha$ is the IMF mismatch parameter defined as $\alpha \equiv M_*^{\rm{LD}}/M_*^{\rm{SPS}}$.}
\label{mstarmeas}
\end{deluxetable}
The stellar mass measured from gravitational lensing and dynamics, $M_*^{\rm{LD}}$, is larger than the masses obtained from the SPS study, $M_*^{\rm{SPS}}$.
This discrepancy can be quantified with the ``IMF mismatch'' parameter $\alpha \equiv M_*^{\rm{LD}}/M_*^{\rm{SPS}}$, also reported in Table \ref{mstarmeas}.
A Salpeter IMF is clearly favored,
while the probability of the IMF being heavier than Chabrier ($\alpha_{\rm{Chab}} >1$) is 95\%.

This result is in agreement with a general trend observed by
\citet{Grillo2009}, \citet{TreuIMF} and \citet{Auger2010L} for the
early-type galaxies of the SLACS sample.  They find that, on average,
a Salpeter IMF better matches the measurements of stellar masses from
lensing and dynamics.  A similar result is found by \citet{Chiara} for
a very massive early-type galaxy.  As discussed extensively by
\citet{TreuIMF}, stellar mass and slope of the dark matter halo are
degenerate with respect to typical lensing and dynamics constraints:
given a bulge-halo decomposition, steepening the dark matter profile
and decreasing the stellar mass can result in fits to the observed
velocity dispersion and mass within the Einstein radius as good as the
original model. \citet{TreuIMF} explained how the observed trend of
increasing $\alpha$ with velocity dispersion can either be interpreted
as the effect of a correlation between IMF or dark matter inner slope
with total mass.  
\citet{Auger2010L} explored this degeneracy by considering adiabatically contracted DM halos set by an imposed relation between stellar and virial mass, and found preference for a stellar mass-to-light ratio closer to a Salpeter than a Chabrier IMF.
Similarly, \citet{NRT} find that a Kroupa IMF, which has a mass-to-light ratio slightly larger than a Chabrier IMF, fits well adiabatically contracted DM halos.
In the present study we allowed the slope of the
dark matter halo of our lens galaxy to vary freely.  Its measured value, $\gamma_{\rm{DM}} =
1.7_{-0.2}^{+0.2}$, is significantly steeper than the inner slope of a
NFW halo.  Still, we find a stellar mass larger than what can be
accounted for with a Salpeter IMF and not compatible with a Chabrier
IMF.  Our results imply that a Salpeter IMF provides a far better
description of the mass-to-light ratio of the stellar population than
a Chabrier IMF even with a steepened dark matter halo.
This result is consistent with the recent findings of \citet{Cap++12} and \citet{v+C11}.

In contrast, Salpeter-like
IMFs are typically ruled out for lower mass systems \citep{Cap++06} or
spiral galaxies \citep{B+d01,Dut++11,Suy++11,Bre++12}.

The lensing and dynamics analysis presented in Sect. \ref{twocompsect}
showed evidence for contraction of the dark matter halo with respect
to a baryonless NFW profile.  
A similar result is found by \citet{Grillo2012} for an ensemble measurement of 39 massive elliptical galaxy lenses.
This result is in qualitative agreement
with many theoretical studies of the evolution of spheroidal galaxies
\citep{Blumenthal,Gnedin04,Gustafsson,Abadi,Duffy}.  \citet{Duffy} in
their simulations of redshift $z=2$ galaxies find inner dark matter
slopes that span the range $1.4 < \gamma_{\rm{DM}} < 2.0$ depending on
the different prescriptions adopted to model the effect of the
baryons.  Our measured value of $\gamma_{\rm{DM}}$ falls nicely in
that range, although our galaxy is at significantly lower redshift.
\citet{Gnedin04} provide a prescription to calculate the dark matter
profile of their modified adiabatic contraction (MAC) model.  It is
interesting to test the MAC model on the measured slope of the dark
matter halo of our galaxy.  The final dark matter density profile of
the MAC model of \citet{Gnedin04} is determined given the observed
light profile, the concentration parameter $c$ of the original (non
contracted) NFW halo and the baryon mass fraction within its virial
radius, $f_b$.

\begin{figure}
\includegraphics[width = \columnwidth]{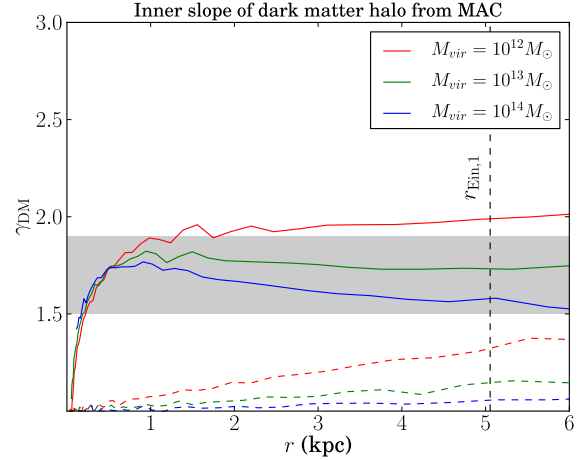}
\caption{{\em Solid lines:} Inner slope of the dark matter halo for modified adiabatic contraction \citep[MAC;][]{Gnedin04} models.
{\em Dashed lines:} Slope of the non-contracted (NFW) dark matter halo.
{\em Shaded region:} 68\% confidence interval of the slope measured in this Paper.}
\label{Gnedinslope}
\end{figure}

Since we do not have information on the initial properties of the dark matter
halo of our galaxy, we use a few trial values of the virial mass
$M_{\rm{vir}}$, spanning a plausible range indicated by a weak lensing
study of ellipticals \citep{Gavazzi07}, and employ a
mass-concentration relation from \citet{Maccio} based on WMAP5
cosmological parameters.  We then calculate the inner slope of the
final dark matter distribution with the software Contra
\citep{Gnedin04}.  The inferred inner slope for
$\log{(M_{\rm{vir}}/M_\Sun)} = 12.0, 13.0, 14.0$ is plotted in
Fig. \ref{Gnedinslope}.  Despite the large range of virial mass
explored, the slopes of the contracted halos lie around $1.5 <
\gamma_{\rm{DM}} < 2.0$ over the spatial region covered by our data.
The MAC model is therefore able to reproduce our measurement of the
dark matter halo slope.

\subsection{A formation scenario}
\label{ssec:form}

As our data show, the stellar distribution in the lens galaxy consists of two components that differ in alignment, surface brightness and stellar population.
This particular structure suggests different formation histories for the two components.
The bright and compact component may have formed first, and later on accreted stellar systems in the outskirts without disrupting the structure of the original bulge.
Alternatively, component 2 might have been present originally and component 1 be formed in a star formation event following a wet merger.
We point out that in the infrared image we see evidence for tidal distortion in the outskirts of the galaxy (see Figure \ref{IRtidal}), possible indication of an ongoing merger. Part of the faint extended envelope of component 2 could be material accreted relatively recently.
The presence of the dust lane in the center of the galaxy (see \S~\ref{ssec:dust}) may also be the result of a recent merger.
We also note that \citet{Vegetti} detected a compact substructure of mass $\sim 3\times10^9M_\Sun$ located in the proximity of the inner ring image, indicating that minor mergers may still be occurring.

Let us consider our first hypothesis: the galaxy consisted initially of the compact component 1.
What are the structural parameters of component 1 and how does it relate to other elliptical galaxies?
Its effective radius is $r_{\rm{eff}} = 0.50''$ (see Tables \ref{TableFits} and \ref{twotNIEfit}), which corresponds to a physical radius of $1.79\mbox{ kpc}$.
Similar effective radii are found for high redshift ($z>1.2$) ellipticals \citep{Daddi,Trujillo,vanDokkum}.
Its stellar mass as inferred from the SPS analysis is given by $\log{(M_*^{\rm{SPS}}/M_\Sun)} = 10.85$ ($\log{(M_*^{\rm{SPS}}/M_\Sun)} = 11.13$) for a Chabrier (Salpeter) IMF.
Local ellipticals with similar values of the stellar mass have effective radii a factor of a few larger than this object \citep{Shen,Hyde}.
Analogously, the high redshift objects of \citet{Daddi}, \citet{Trujillo} and \citet{vanDokkum} are also significantly more massive than local galaxies with similar effective radii.
Finding objects in the local universe that correspond to these high redshift ``red nuggets'' is in fact a standing problem in the study of elliptical galaxies.
It is not clear how objects initially so compact evolve into the more diffuse galaxies that we observe at recent times.

Recent numerical simulations \citep{Hopkins2009,Oser} showed how minor
dry mergers can increase the size of elliptical galaxies
significantly, with the stars of the accreted objects that grow the
outskirts of the galaxy, even though the observed and predicted merger
rates are such that this mechanism might not be sufficient
\citep{New++11a}. The observational signature of this process would be
the presence of a compact core, the original red nugget, surrounded by
a more diffuse distribution of stars from the accreted systems.  The
galaxy studied in this paper might be one of these objects.

\section{Summary}

\begin{figure}
\includegraphics[width = \columnwidth]{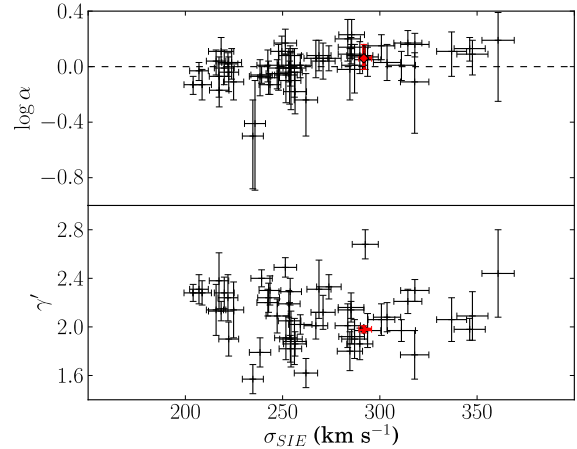}
\caption{{\em Top panel:} IMF mismatch parameter $\alpha \equiv M_*^{\rm{LD}}/M_*^{\rm{SPS}}$ relative to a Salpeter IMF vs. lens strength $\sigma_{\rm{SIE}}$ for the SLACS lenses of \citet{TreuIMF} (black crosses) and for the Jackpot (red cross). Measurements of \citet{TreuIMF} are obtained assuming a NFW dark matter halo with fixed scale radius for the lensing and dynamics analysis.
{\em Bottom panel:} average slope of the total density profile $\gamma'$ vs. lens strength $\sigma_{\rm{SIE}}$ for the SLACS lenses of \citet{PaperX} (black crosses) and for the Jackpot (red cross).}
\label{comparison}
\end{figure}

We have presented a new set of photometric and spectroscopic data for the gravitational lens system SDSSJ0946+1006.
We used these data to constrain the structural properties of the foreground elliptical galaxy of the system.
On the basis of our results, the following statements can be made.
\begin{itemize}

\item The redshift of the source corresponding to the outer ring is $z_{s2} = 2.41_{-0.21}^{+0.04}$ at 68\% confidence level, as revealed by our photo-z measurement.
\item If we describe the total mass distribution with a power-law
ellipsoid $\rho \propto r^{-\gamma'}$, lensing and dynamics data give
as measured value $\gamma' = 1.98\pm 0.02\pm0.01$.  This parameter
should be interpreted as an effective slope of the density profile
averaged over the region within the outer Einstein ring.  The special
lensing configuration and the exquisite data quality of our data,
allowed us to measure $\gamma'$ with unprecedented precision.  The
value obtained is consistent with isothermal ($\gamma' = 2$) and is in
agreement with the general trend observed for the massive early-type
galaxies of the SLACS sample, $\left<\gamma'\right> = 2.078\pm0.027$
with intrinsic scatter $\sigma_{\gamma'} = 0.16\pm0.02$
\citep{PaperX,Koo++09,Bar++11}.  See Figure \ref{comparison} for a comparison of our
measurement of $\gamma'$ with measurements of the same parameter for
the SLACS sample of early-type galaxies by \citet{PaperX}.
\item We are able to decompose dark and stellar matter with lensing and dynamics data, assuming a power-law density profile for the dark matter.
The derived stellar mass is $5.5_{-1.3}^{+0.4}\times10^{11}M_\Sun$, consistent with a Salpeter IMF and inconsistent with a Chabrier IMF.
This constraint on the IMF is plotted in Figure \ref{comparison} together with similar measurements for the other SLACS lenses obtained by \citet{TreuIMF}.
Note that we achieve better precision despite using less strict assumptions on the dark matter profile.
\item The slope of the dark matter halo is found to be
$\gamma_{\rm{DM}} = 1.7\pm0.2$.  This is a strong evidence for
contraction relative to the $r^{-1}$ behavior of NFW profile observed
in simulations without baryons, and is in agreement with the inner
dark matter profiles obtained by \citet{Duffy} in their simulations of
$z=2$ galaxies and with the MAC model of \citet{Gnedin04}. 
Our inferred bulge-halo decomposition has a local projected slope at the inner ring
in agreement with the value measured by \cite{Vegetti} based on a completely independent technique.
\item The particular structure of the stellar distribution, with a
compact core and a misaligned faint extended envelope, might be the
result of accretion of low mass systems by a compact red nugget.
\item A spectroscopic detection of the redshift of the outer ring would still help improve the model, but would not lead to a dramatic change in the results of our analysis.
\end{itemize}

\begin{acknowledgments}
Based on observations made with the NASA/ESA Hubble Space Telescope, obtained at the Space Telescope Science Institute, which is operated by the Association of Universities for Research in Astronomy, Inc., under NASA contract NAS 5-26555. These observations are associated with programs 11701, 11202, and 10886. Support for those programs was provided by NASA through a grant from the Space Telescope Science Institute, which is operated by the Association of Universities for Research in Astronomy, Inc., under NASA contract NAS 5-26555.

Some of the data presented herein were obtained at the W.M. Keck Observatory, which is operated as a scientific partnership among the California Institute of Technology, the University of California and the National Aeronautics and Space Administration. The Observatory was made possible by the generous financial support of the W.M. Keck Foundation
The authors wish to recognize and acknowledge the very significant cultural role and reverence that the summit of Mauna Kea has always had within the indigenous Hawaiian community.  We are most fortunate to have the opportunity to conduct observations from this mountain.

T.~Treu acknowledges support from the Packard Foundation through a Packard Research Fellowship.
R.~Gavazzi acknowledges support from the Centre National des Etudes Spatiales. P.~J.~Marshall was given support by the Royal Society in the form of a research fellowship.
\end{acknowledgments}

\bibliography{Jackpot_refresp}{}
\bibliographystyle{apj}


\appendix

\section{Dust correction}\label{ssec:dust}

The presence of dust complicates our analysis. Nevertheless, we deal with it by applying a procedure similar to that adopted by \citet{Koop2003} and \citet{Suyu2009} for the system B1608+656. The details of the procedure are the following.

We select a small region in the center of the galaxy for which we want
to apply a dust correction.  We estimate the intrinsic colors of this
central part by measuring them in a region that we think is free of dust.

We assume a dust law from \citet{Cardelli}, with $R_V = 3.1$.  Given
the flux in one band and the colors of the object, the fluxes in the
remaining bands are determined by the dust law.  Therefore, with
images in two or more bands we can constrain both the intrinsic flux
and the dust content of the object.  In our case we determine these
two quantities in the central region of the lens on a pixel-by-pixel
basis by fitting the F814W, F606W and F438W fluxes.  We want to check
if we can account for the dark lane observed in the F336W image
independently from the data in that band, therefore we do not include
that image in our dust analysis. The F160W image is excluded because
of its lower spatial resolution.  The PSF of the different images is
not matched. Our inference on the presence of dust is not
affected by this approximation.

\begin{figure*}[!]
  \begin{center}
    \begin{tabular}{ccc}
      \includegraphics[width=0.33\textwidth]{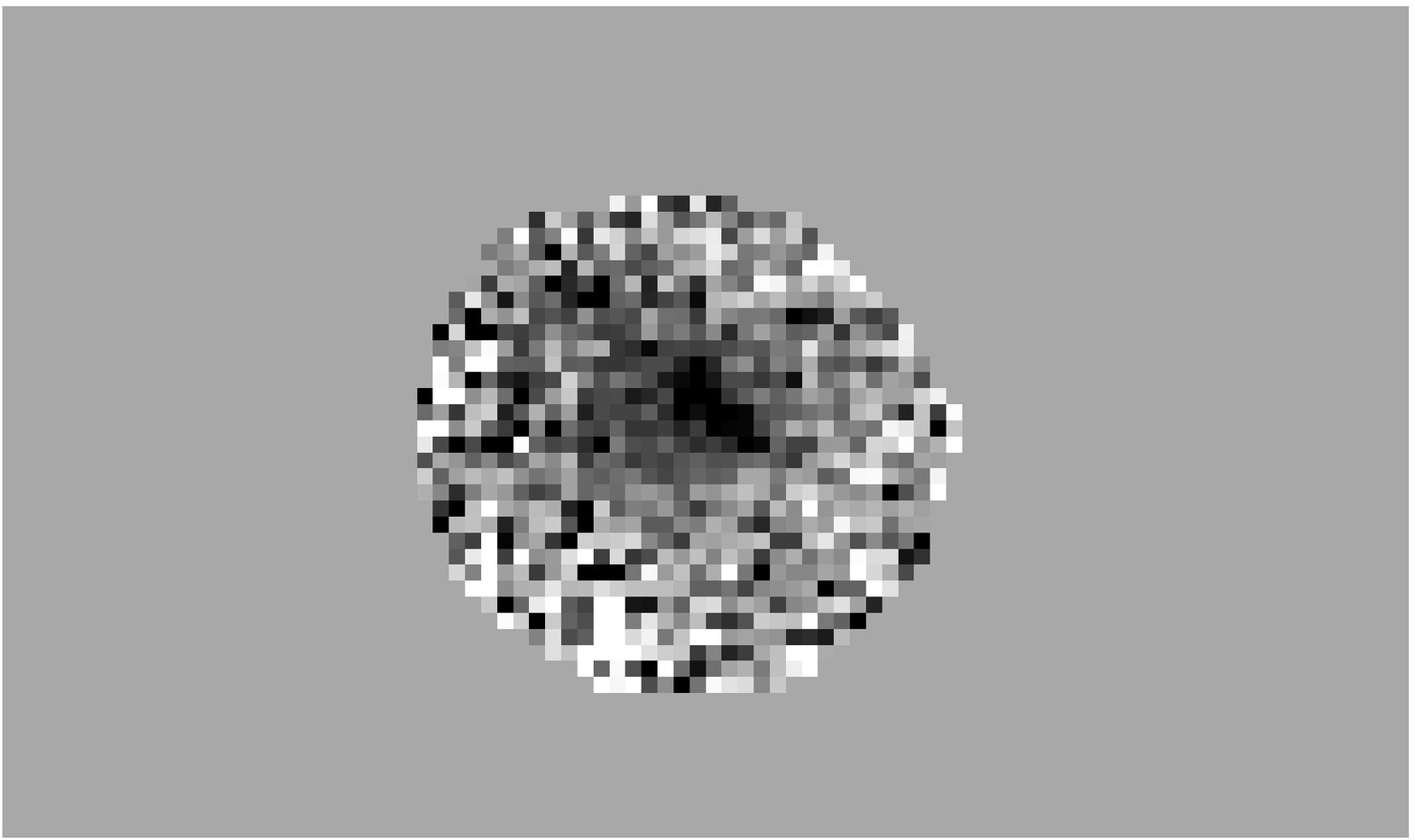} &
      \includegraphics[width=0.33\textwidth]{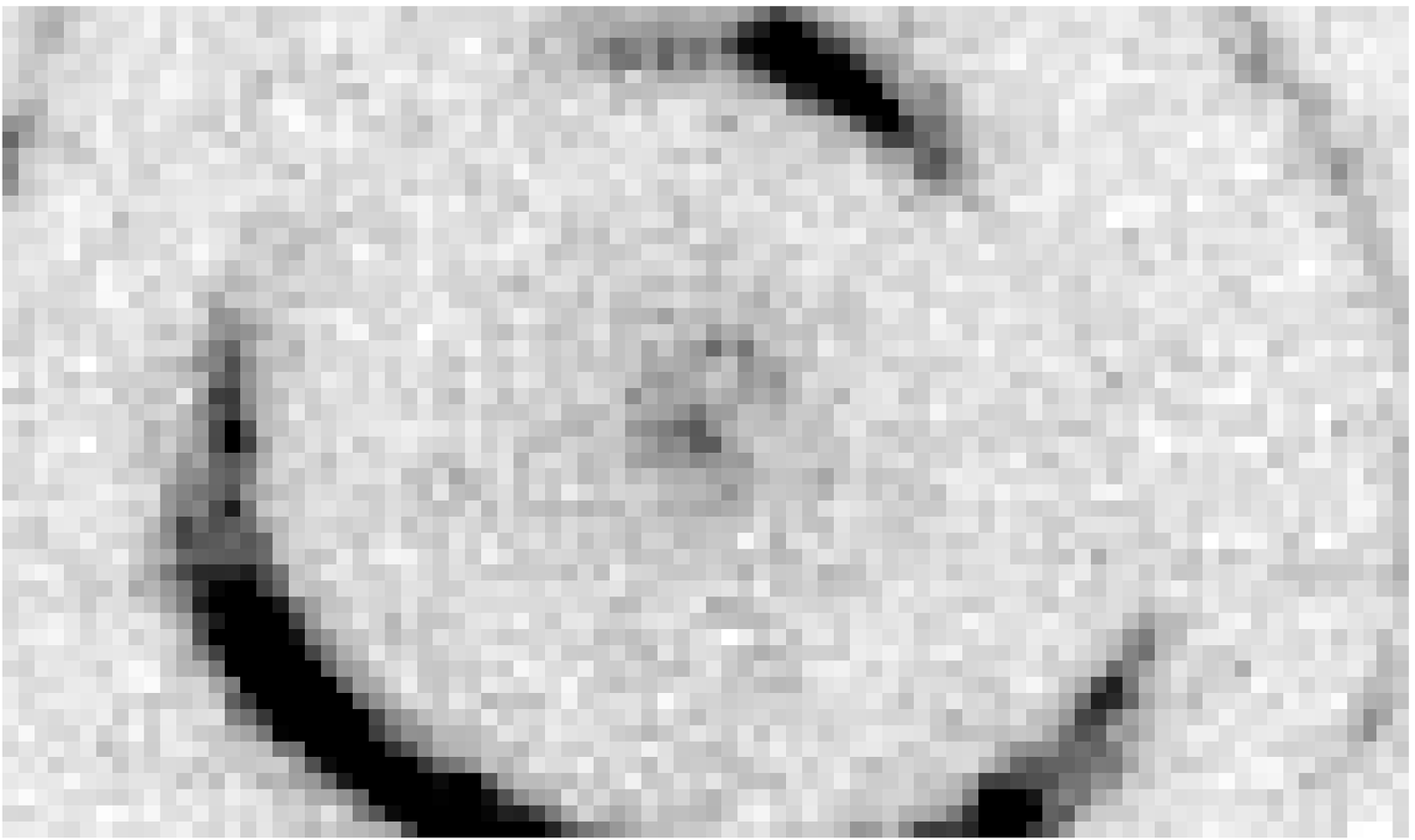} &
      \includegraphics[width=0.33\textwidth]{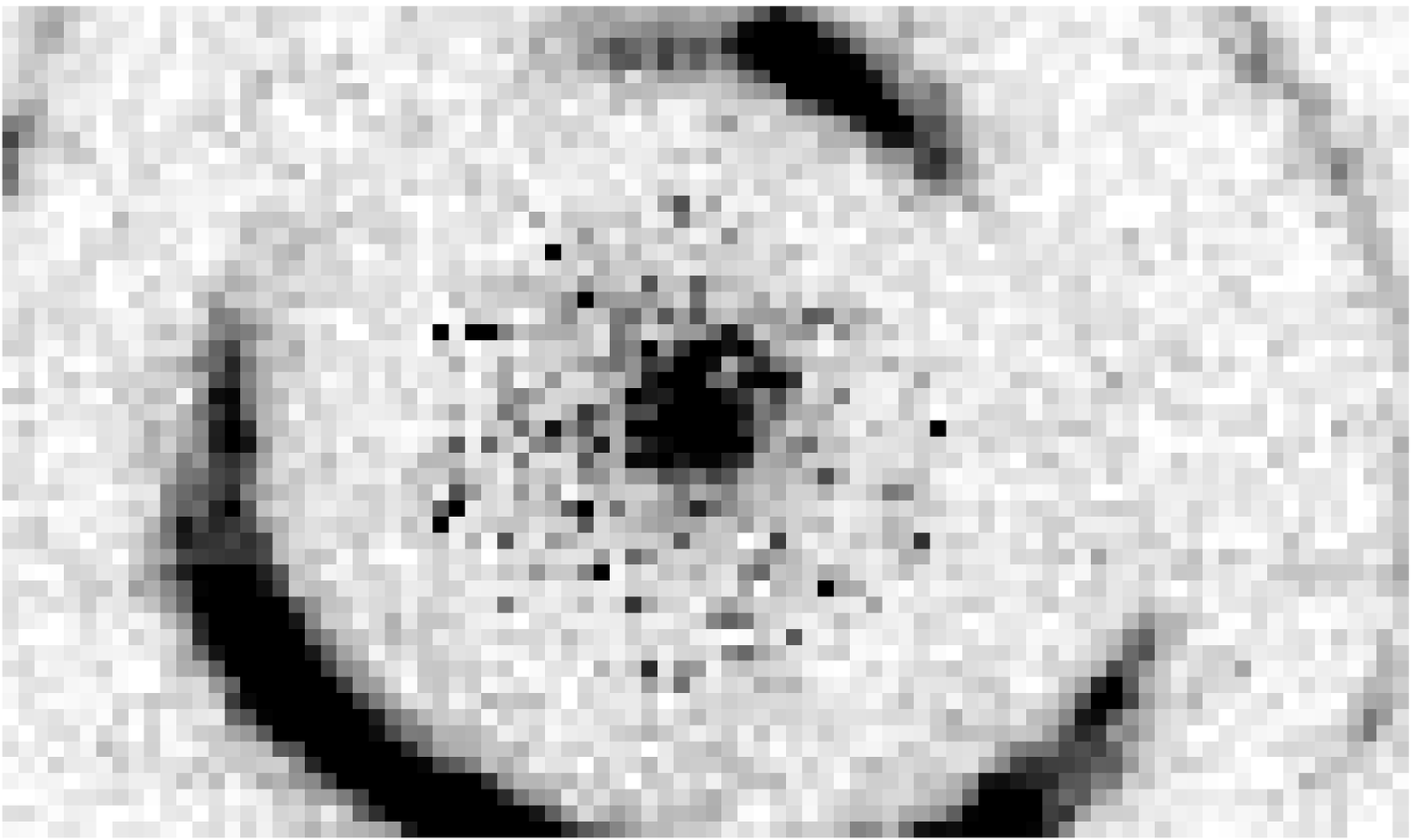}
    \end{tabular}
    \caption{Dust correction. {\em Left panel:} Dust map of the central part of the galaxy based on F438W, F606W and F814W images. Dark pixels represent higher column densities of dust. {\em Middle panel:}  Original image in the F336W band. {\em Right panel:} Dust-corrected F336W band image, showing a single clump of light. The central parts have lower signal-to-noise ratio as a result of the noisy dust map.}
    \label{Dustmap}
  \end{center}
\end{figure*}

The two dust-free colors, F606W - F814W and F438 - F606W, are measured in an annulus around the center and inside of the inner ring.
We cannot rule out the presence of a uniform dust screen, but that would not affect our conclusions as the tools that we use for quantitative analyses can account for that.
The dust-corrected flux in the lens center is then calculated with the fitting method described in Sect. 5.4 of \citet{Suyu2009}.
Figure \ref{Dustmap} shows the recovered dust map,
the F336W image corrected for dust and its original version.  It can
clearly be seen how the dust map, obtained without using data from the
F336W, has largest column density right where we observe the dark lane
in the image. In the dust-corrected image, the lens looks indeed more
like a single object.
The overall amount of dust is relatively small, as the correction to the total magnitude in the F606W band is approximately $0.10$ mags.

\section{Ellipticity effects}

The mass and light distributions of the lens galaxy are well approximated by ellipses, but we make the assumption of spherical symmetry for the analysis of the kinematics data.
How does this assumption affect the results we present?
To answer this question we make use of the axisymmetric version of the Jeans equations \citep{BinTrem}
\begin{eqnarray}
\bar{v_R^2}(R,z) = \bar{v_z^2}(R,z) &=& \frac{1}{\rho_*(R,z)}\int_z^{\infty}dz'\rho_*(R,z')\frac{\partial\Phi}{\partial z'} \\
\bar{v_\phi^2}(R,z) &=& \bar{v_R^2} + \frac{R}{\rho_*}\frac{\partial(\rho_8\bar{v_R^2})}{\partial R} + R\frac{\partial\Phi}{\partial R},
\end{eqnarray}
obtained assuming a distribution of stars of the form $f(E,L_z)$.

\begin{figure}
\begin{tabular}{cc}
\includegraphics[width=0.5\textwidth]{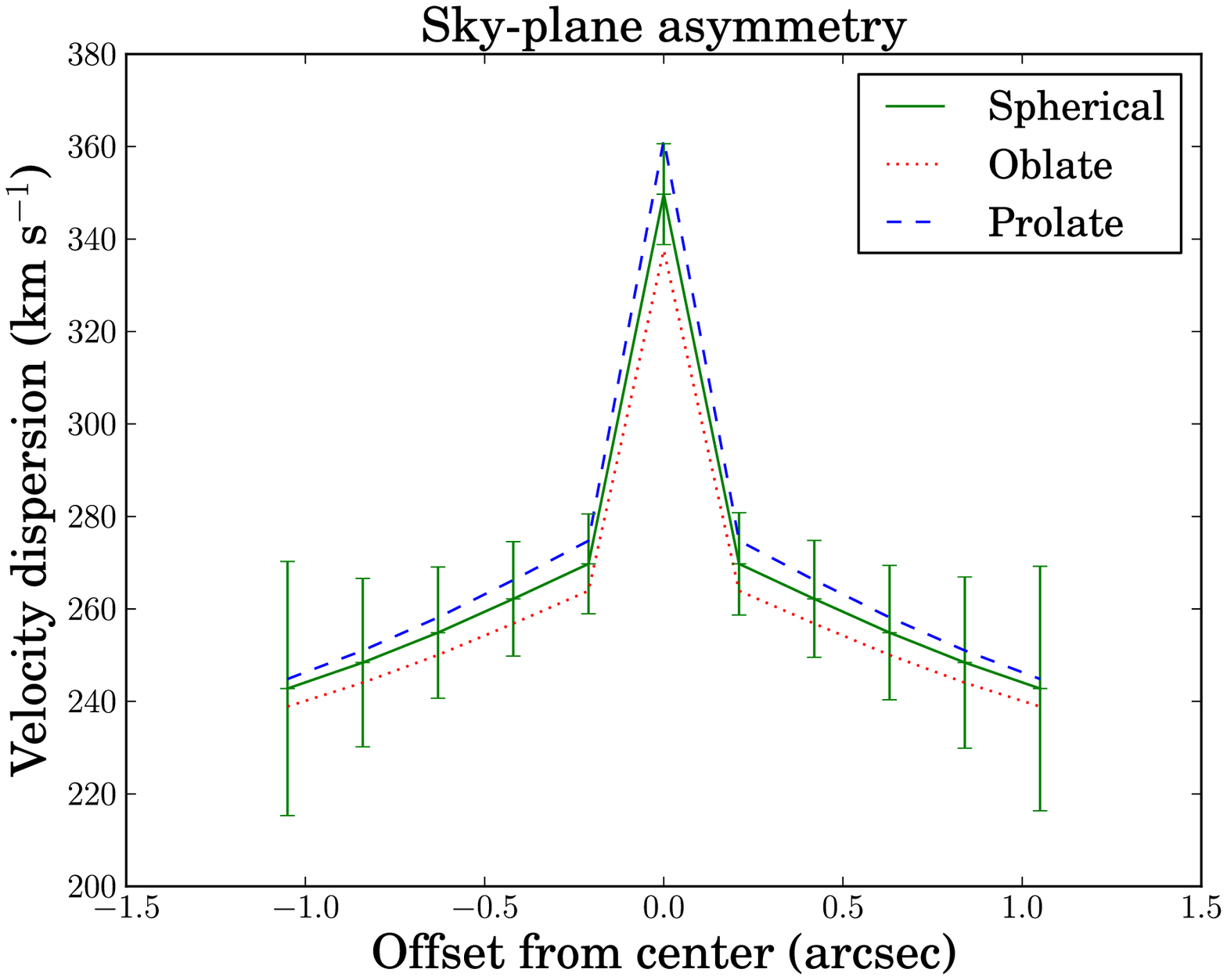} &
\includegraphics[width=0.5\textwidth]{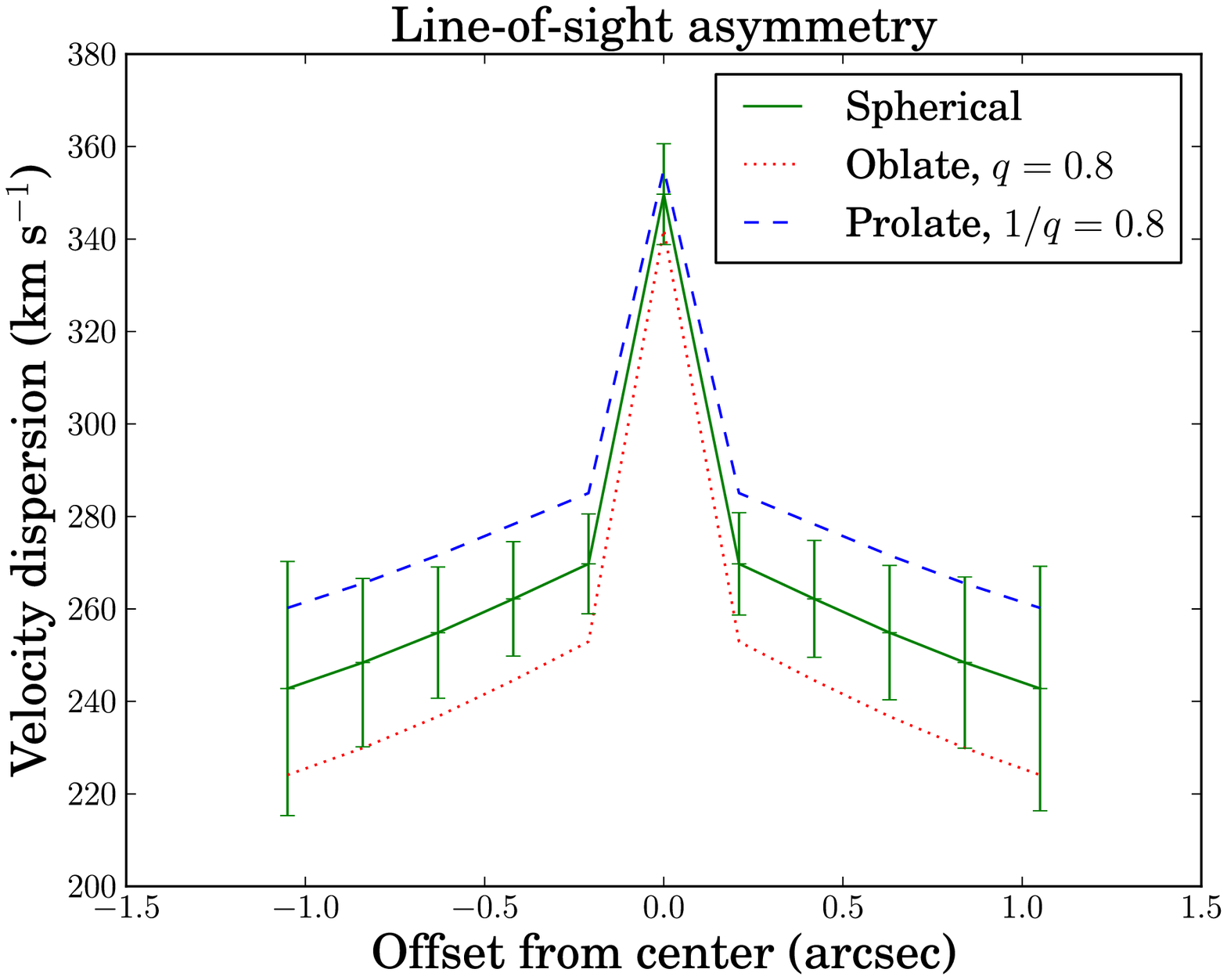} \\
\end{tabular}
\label{ellipt1}
\caption{{\em Top:} 
Line-of-sight velocity dispersion profile for a spherical model and for elliptical models with $q_* = 0.95$ and $q = 0.87$, calculated by solving the axisymmetric Jeans equation. The models have the same projected mass within the inner Einstein radius.
PSF smearing is not included, resulting in the high central peak.
Overplotted are the error bars on the measured velocity dispersion profile.
{\em Bottom:} Velocity dispersion profile of the spherical model and of an oblate (prolate) ellipsoid with minor (major) axis parallel to the line of sight and axis ratio 0.8.}
\end{figure}

We take our best-fit single component model from Section \ref{Jeanssect} and make it elliptical by assuming that the rotation axis $L_z$ is in the plane of the sky and fixing the projected ellipticities in light and mass to $q_*=0.95$ and $q = 0.87$ respectively.
The first value is the ellipticity of the best single-component fit to the light profile, the latter value is given in \citet{Gavazzi}.
Then we assume isotropy in the velocity dispersion tensor, calculate the line of sight velocity dispersion profile in the two possible cases of oblate or prolate ellipsoid, and compare it to the corresponding spherical case.
Results are shown in Fig. \ref{ellipt1}.
Deviations from spherical symmetry bring differences on the order of a few km s$^{-1}$ on the velocity dispersion profile, well within our uncertainties on the measurements, and therefore is not a concern for possible biases.

More important are the effects of asymmetries along the line of sight.
We do not have any direct measurement of the line-of-sight structure of the lens, but from the observed projected flattening we can get information on the intrinsic shape of the galaxy by statistical means.
\citet{Padilla} measured the distribution of intrinsic axis ratios of massive elliptical galaxies.
By drawing samples of galaxy shapes from their inferred distribution and assuming random orientations we find that $68\%$ of the objects that produce a projected ellipticity $q_*=0.95$ have an axis ratio rounder than $0.8$.
How does the velocity dispersion profile of an oblate (prolate) galaxy with minor (major) axis along the line of sight and axis ratio of $0.8$ differ from that of a spherical galaxy with the same (observed) projected mass within the Einstein radius?
We use the axisymmetric Jeans equation to address this question as well.
We take our best-fit spherical model and modify it into an oblate (prolate) ellipsoid with the axis ratio of both the light and mass distribution fixed at $0.8$, orienting $L_z$ along the line of sight.
The line-of-sight velocity dispersion profile for isotropic orbits in the oblate and prolate case is also plotted in Fig. \ref{ellipt1}.
The spread relative to the spherical case is somewhat larger than the uncertainties.
To make sure that our assumption of spherical symmetry does not alter the measurements presented in this Paper we recalculate the inference of the key model parameters by inflating the error bars on the velocity dispersion measurements by a factor $1.5$, matching the scatter introduced by the unknown line-of-sight oblateness or prolateness of the lens.
None of the results changes appreciably.
The lens modeling does not depend on assumptions on the line-of-sight mass distribution,
and so in this regard our results are robust.

\end{document}